\documentclass{aa}

\usepackage{graphicx}
\usepackage{txfonts}
\usepackage{lipsum}
\usepackage{subcaption}         
\usepackage{lscape}             
\usepackage{placeins}           
\usepackage{xcolor}
\usepackage{wasysym}

\def\chisqr{\hbox{$\chi^2_{\rm r}$}}
\def\msun{\hbox{${\rm M}_{\odot}$}}
\def\mjup{\hbox{${\rm M}_{\jupiter}$}}

\def\mspy{\hbox{${\rm M}_{\odot}$\,yr$^{-1}$}}

\def\rsun{\hbox{${\rm R}_{\odot}$}}

\def\rcor{\hbox{$r_{\rm cor}$}}
\def\rmag{\hbox{$r_{\rm mag}$}}
\def\mstar{\hbox{$M_{\star}$}}
\def\rstar{\hbox{$R_{\star}$}}

\def\teff{\hbox{$T_{\rm eff}$}}
\def\logg{\hbox{$\log g$}}

\def\vD{\hbox{$v_{\rm D}$}}

\def\ms{\hbox{m\,s$^{-1}$}}

\def\kms{\hbox{km\,s$^{-1}$}}

\def\vsini{\hbox{$v \sin i$}}

\def\mic{\hbox{$\mu$m}}

\def\emr{}
\def\Bl{\hbox{$B_{\rm \ell}$}}
\def\Bd{\hbox{$B_{\rm d}$}}
\def\Bo{\hbox{$B_{\rm o}$}}

\def\degr{\hbox{$^\circ$}}

\def\Mdot{\hbox{$\dot{M}$}}

\def\Prot{\hbox{$P_{\rm rot}$}}

\newcommand{\hei}{\hbox{He$\;${\sc i}}}

\newcommand{\pab}{\hbox{Pa${\beta}$}}
\newcommand{\brg}{\hbox{Br${\gamma}$}}

\begin{document}


\title{Unstable magnetospheric accretion on the T~Tauri star TW~Hya} 

   \author{J.-F.~Donati\inst{1}
      \and P.I.~Cristofari\inst{2}
      \and C.~Moutou\inst{1}
      \and A.~Carmona\inst{1}
      \and A.~Lavail\inst{1}
      \and J.~Bouvier\inst{3}
      \and K.~Perraut\inst{3}
      \and S.H.P.~Alencar\inst{4} 
      \and M.~Audard\inst{5}
      \and K.~Grankin\inst{6}
      \and F.~M\'enard\inst{3} 
      \and M.~Takami\inst{7}
      \and the SPIRou science team
          }
   \institute{Univ.\ de Toulouse, CNRS, IRAP, 14 avenue Belin, 31400 Toulouse, France (\email{jean-francois.donati@cnrs.fr})
        \and Leiden Observatory, Leiden University, Niels Bohrweg 2, 2333 CA Leiden, the Netherlands
	\and Univ.\ Grenoble Alpes, CNRS, IPAG, 38000 Grenoble, France 
        \and Departamento de F\'{\i}sica -- ICEx -- UFMG, Av. Ant\^onio Carlos, 6627, 30270-901 Belo Horizonte, MG, Brazil
        \and Department of Astronomy, University of Geneva, Chemin Pegasi, 51, Versoix CH-1290, Switzerland 
        \and Crimean Astrophysical Observatory, Nauchny, Crimea 298409 
        \and Institute of Astronomy and Astrophysics, Academia Sinica, Roosevelt Rd, Taipei 10617, Taiwan 
             }

\date{Submitted 2026 xxx -- Accepted 2026 yyy} 

\abstract{
In this paper we present new spectropolarimetric and velocimetric observations of the prototypical classical T~Tauri star TW~Hya obtained with 
SPIRou at the Canada-France-Hawaii Telescope, expanding our previous monitoring over two new seasons (2024 and 2025).  We confirm that the 
large-scale magnetic field of TW~Hya varied with time, and find that it showed fluctuations on a timescale of about a year in addition to the 
longer term variations outlined in the previous study.  Using Zeeman-Doppler imaging, we obtain that the large-scale field of TW~Hya mostly 
consisted of a poloidal dipole of mean polar strength 0.83~kG, inclined at an average 17\degr\ to the rotation axis.  We also find that the 
radial velocities of TW~Hya, once fully filtered from telluric contamination, were dominated by rotational modulation induced by activity, with 
residuals of 32~\ms\ rms.  No signal from a putative close-in planet is found, with an upper limit on the planet mass ranging from 
0.33 to 0.98~\mjup\ for distances of 0.053 to 0.41~au from the central star.  Emission lines indicate that the mass accretion rate 
was equal to $10^{-8.33\pm0.20}$~\mspy\ on average, with peak-to-peak fluctuations by a factor of $\simeq$5 from season to season.  This confirms 
that accretion onto TW~Hya is unstable, with the magnetospheric gap carved by the large-scale field at the center of the disk extending on average 
no further than 33$-$40\% of the corotation radius where the disk Keplerian angular velocity equals the rotation rate at the stellar surface.  
}


\maketitle



\section{Introduction}
\label{sec:int}

Low-mass stars form from collapsing turbulent giant molecular clouds evolving into flat rotating structures called accretion disks 
as they contract.  Protostars are born in and feed from the central regions of these flat structures, while protoplanetary systems progressively assemble 
in the surrounding areas.  Magnetic fields and turbulence, in addition to gravity, are known to play key roles in the process, with the fields being instrumental 
to ensure evacuation of the initial angular momentum content of the parent cloud through magnetic braking via jets and outflows \citep{Andre14,Lebreuilly24}.  
At an age of a few Myr, once low-mass protostars have emerged from their dust cocoons, they are called pre-main-sequence (PMS) T~Tauri stars, and more 
specifically classical T~Tauri stars (cTTSs) when they still accrete material from their disks.  They are ideal objects for studying the late 
phases of star formation and the complex magneto-hydro-dynamical (MHD) mechanisms involved in this process \citep[e.g.,][]{Romanova21}, and for 
constraining models of planetary formation and migration at a stage for which little observational material yet exists.  In this task, one needs to 
observe cTTSs using spectroscopy to investigate accretion / ejection processes taking place in the central regions of their accretion disks 
\citep[e.g.,][]{Bouvier07}, but also spectropolarimetry to detect and reconstruct the large-scale topologies of their magnetic fields 
\citep[e.g.,][]{Donati09} and velocimetry to investigate the potential presence of close-in planets \citep[e.g.,][]{Crockett12}.  Moreover, such 
monitoring observations must be repeated regularly, including in the long term, to follow the evolution of magnetic topologies \citep{Donati24b}, of 
accretion patterns \citep{Fischer23}, and of radial velocity (RV) variations caused by putative close-in planets and hidden within the activity jitter 
\citep[e.g.,][]{Zaire24}.  

At a distance of only 60~pc, TW~Hydrae is the cTTS closest to us and a prototypical PMS star of obvious interest for studies of forming 
low-mass stars and their planetary systems.  Already observed quite extensively, TW~Hya is well characterized, with a mass of $0.80\pm0.05$~\msun\ 
\citep[in agreement with the dynamical mass inferred with ALMA,][]{Teague19}, a radius of $1.16\pm0.13$~\rsun\ and an age of about 8~Myr 
\citep[see][for a discussion of these parameters]{Donati24b}.  In particular, its magnetic field was clearly detected, both through polarization in 
spectral lines and Zeeman broadening of unpolarized lines \citep{Yang07,Donati11b,Sokal18} and found to be variable with time \citep{Donati24b}.  
The accretion properties of TW~Hya were monitored over the past 25~yr \citep{Herczeg23,Ji26}, demonstrating that the central star has been accreting 
from its disk at variable rates over this period.  

TW~Hya features a highly structured transition disk featuring several gaps, one of which around 1~au 
\citep{Andrews16,vanBoekel17}, with the regions within 4~au containing optically thin dust only while the outer regions feature optically thick 
submicron-sized grains \citep{Calvet02,Henning24}. TW~Hya is thus an ideal candidate to study planet formation and dust dynamics, and to search for potential 
protoplanets in the disk that may have carved the reported gaps and triggered dust filtering.  Attempts at directly detecting such planets, whose masses were 
estimated to be of order 0.1~\mjup\ on the basis of the disk structure, have been inconclusive so far \citep[e.g.,][]{Huelamo22}.  
TW~Hya was also scrutinized in velocimetry to search for potential close-in planets \citep{Setiawan08,Huelamo08,Donati11b,Donati24b}, with the conclusion that 
its RV curve is dominated by activity and that any close-in planet, if present, must be less massive than a fraction of \mjup.  

With this new study, we expanded the monitoring of TW~Hya with the near-infrared (nIR) SPIRou spectropolarimeter / velocimeter \citep{Donati20} at the Canada-France-Hawaii 
Telescope (CFHT) that we initiated in 2019 and pursued until 2022 \citep{Donati24b}, adding observations in 2024 and 2025 that doubled the total number 
of visits.  After detailing the full data set in Sec.~\ref{sec:obs}, we investigated the magnetic variability over the full 
campaign in Sec.~\ref{sec:mag} and used Zeeman-Doppler imaging (ZDI) to achieve a consistent modeling of the large-scale and small-scale magnetic field 
of TW~Hya for each observed season (see Sec.~\ref{sec:zdi}).  We similarly revisited the analysis of the RV data in Sec.~\ref{sec:rvs} and of the accretion 
properties of TW~Hya in Sec.~\ref{sec:act}.  We finally summarized our main results in Sec.~\ref{sec:dis}.

\section{SPIRou observations}
\label{sec:obs}

SPIRou records unpolarized and polarized stellar spectra over the 0.95--2.50~\mic\ spectral range at a resolving power of 70,000 in a single exposure. 
As in our previous study \citep{Donati24b}, we focus on the unpolarized (Stokes $I$) and circularly polarized (Stokes $V$) spectra.  Polarimetric observations are 
acquired as sequences of four subexposures, each obtained with a different azimuth of the Fresnel-rhomb retarders in the SPIRou polarimetric module.  This observing 
strategy suppresses instrumental polarization systematics to first order \citep{Donati97b}.  Each sequence yields one Stokes $I$ spectrum, one Stokes $V$ spectrum, and 
a diagnostic null-polarization spectrum ($N$) used to identify possible instrumental or data-reduction artefacts.  In addition to the 84 SPIRou spectra collected 
in our previous study, of which 82 were retained for analysis, we acquired 84 further spectra as part of the SPICE Large Program with SPIRou at CFHT (RUNIDs 24AP46 and 
25AP45, PI J.-F.~Donati).  These comprise 38 observations obtained from February to May 2024 and 46 from February to June 2025.   
One polarimetric sequence was acquired per night, except on a few nights affected by poor weather when two consecutive sequences were obtained.  Two of the new 
spectra, recorded on 2025 March~19 and June~11, were discarded because of their low signal-to-noise ratios (S/N), leaving 82 new spectra suitable for analysis. 
The final data set thus consists of 164 usable spectra with H-band S/N ranging from 140 to 410 (median 340).  The complete log of 
observations, spanning 2248~d, is presented in Table~\ref{tab:log}.  Rotational phases and cycle numbers were computed using a rotation period of 3.587~d (see 
Sec.~\ref{sec:mag}), adopting a reference barycentric Julian date of 2458488.5 \citep[as in][]{Donati24b}.

All spectra were reduced with \texttt{Libre ESpRIT}, the standard ESPaDOnS reduction pipeline at CFHT, optimized for spectropolarimetric observations and adapted to 
SPIRou \citep{Donati20}.  Least-Squares Deconvolution \citep[LSD;][]{Donati97b} was then applied to the reduced spectra using a line mask generated from the VALD-3 
database \citep{Ryabchikova15} for atmospheric parameters appropriate to TW~Hya ($\teff=3750$~K, $\logg=4.5$; see Sec.~\ref{sec:mag}).  The mask includes only atomic lines 
deeper than 10\% of the continuum level $I_c$ and comprises about 1500 lines with a mean wavelength of 1750~nm and an effective Land\'e factor of 1.2.
The resulting Stokes $V$ LSD profiles have noise levels, $\sigma_V$, ranging from 1.6 to 6.4 (median 2.0) in units of $10^{-4},I_c$.  Clear Zeeman signatures are detected 
in nearly all observations, with typical peak-to-peak amplitudes of about 0.1\%.  The corresponding longitudinal magnetic field, \Bl\ (the line-of-sight component of the magnetic 
field averaged over the visible stellar hemisphere), was derived from the LSD Stokes $I$ and $V$ profiles following \citet{Donati97b}, integrating over a velocity interval 
of $\pm40$~\kms, appropriate for TW~Hya.  Assuming $\Bl=0$ yields a reduced chi-square of $\chisqr=82$, demonstrating a highly significant detection of the longitudinal 
field.  Performing the same analysis on $N$ gives $\chisqr=1.08$, consistent with the absence of spurious polarization signals.  
The measured \Bl\ also exhibits a modest rotational modulation (see Sec.~\ref{sec:mag}).
To investigate the behaviour of magnetically insensitive spectral lines, we also computed Stokes $I$ LSD profiles using a dedicated mask containing only CO bandhead lines 
between 2.29 and 2.40~\mic\ and deeper than 10\% of $I_c$.  This mask comprises about 500 lines and allows a direct comparison between the LSD profiles of 
atomic and CO lines.

We also processed the SPIRou data with the latest version of \texttt{APERO} (v0.7.294), the standard SPIRou reduction pipeline \citep{Cook22} optimized for high 
RV precision and accurate telluric-line correction.  The reduced spectra, corrected for residual spectrograph drifts using the simultaneous Fabry-Perot reference spectrum 
\citep{Donati20}, were analysed with the line-by-line (LBL) technique \citep[v0.65;][]{Artigau22}.  This analysis provides nightly RVs and differential temperature 
measurements, $dT$, derived from variations in the relative depths of spectral lines with respect to their median profiles \citep{Artigau24}.  We find that the LBL RVs 
exhibit clear rotational modulation whereas $dT$ remains essentially unmodulated.  The RVs also show a residual dependence on the barycentric Earth radial 
velocity (BERV; see Sec.~\ref{sec:rvs}).


\section{Magnetic field and temperature changes}
\label{sec:mag}

We found that \Bl\ ranged from $-212$ to 56~G (median error 11~G) and varied significantly, both from season to season and within individual seasons (especially in 2024 and 2025).  
To quantify these temporal fluctuations, we arranged \Bl\ values in a vector denoted $y$ and applied GPR \citep[][]{Haywood14,Rajpaul15}, employing 
the following quasi-periodic (QP) covariance function $c(t,t')$
\begin{eqnarray}
c(t,t') = \theta_1^2 \exp \left( -\frac{(t-t')^2}{2 \theta_3^2} -\frac{\sin^2 \left( \frac{\pi (t-t')}{\theta_2} \right)}{2 \theta_4^2} \right) 
\label{eq:covar}
\end{eqnarray}
where $\theta_1$ is the amplitude (in G) of the Gaussian process (GP), $\theta_2$ is its recurrence period (measuring \Prot), $\theta_3$ is the evolution timescale on which the \Bl\ curve 
changes shape (in d), and $\theta_4$ is a smoothing parameter controlling the amount of harmonic complexity.
A fifth hyperparameter, $\theta_5$, describes the excess of uncorrelated noise on \Bl.  This ensures that the QP GPR fit can diagnose cases where the standard error 
bars are underestimated, and reach the solution with highest likelihood, $\mathcal{L}$, defined by
\begin{eqnarray}
2 \log \mathcal{L} = -n \log(2\pi) - \log|C+\Sigma+S| - y^T (C+\Sigma+S)^{-1} y
\label{eq:llik}
\end{eqnarray}
where $C$ is the covariance matrix for all observing epochs, $\Sigma$ is the diagonal variance matrix associated with $y$, $S=\theta_5^2 J$ is the contribution of the additional
white noise, with $J$ the identity matrix, and $n$ is the number of data points.

\begin{table}[t!]
\caption{MCMC GPR modeling of the \Bl\ curve of TW~Hya}
\centering
\resizebox{\linewidth}{!}{
\begin{tabular}{cccc}
\hline
Parameter   & Symbol & Value & Prior   \\
\hline
GP amplitude (G)     & $\theta_1$  & $75\pm12$        & mod Jeffreys ($\sigma_{\Bl}$) \\
Rec.\ period (d)     & $\theta_2$  & $3.587\pm0.009$  & Gaussian (3.6, 0.5) \\
Evol.\ timescale (d) & $\theta_3$  & $108\pm24$       & log Gaussian ($\log$ 100, $\log$ 2.0) \\ 
Smoothing            & $\theta_4$  & 2.8              & fixed   \\
White noise (G)      & $\theta_5$  & $10.9\pm1.3$     & mod Jeffreys ($\sigma_{\Bl}$) \\
Rms (G)              &             & 14.0             & \\ 
$\chisqr$            &             & 1.57             & \\
\hline
\end{tabular}}
\tablefoot{For each hyperparameter, we list the fitted value along with the corresponding error bar and the assumed prior.  The knee of the modified Jeffreys prior is set to
the median error of \Bl\ (11~G).  The smoothing factor $\theta_4$, poorly constrained by the data, was set to 2.8 (the optimum in a preliminary unconstrained run).  }
\label{tab:gpb}
\end{table}

\begin{figure}[ht!]
\centerline{\includegraphics[scale=0.38,angle=-90]{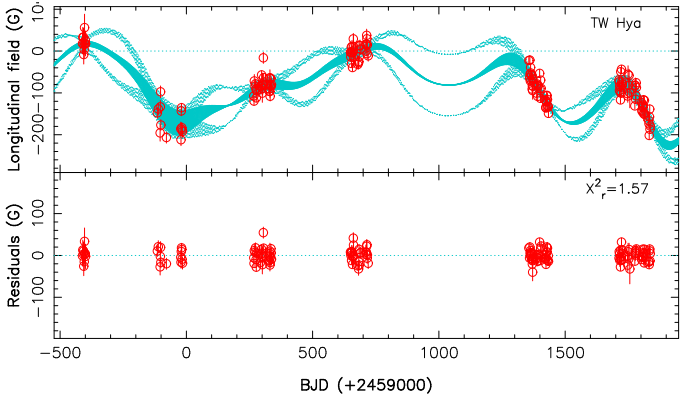}\vspace{2mm}}
\caption[]{Longitudinal magnetic field \Bl\ (red dots), and QP GPR fit to the data (full cyan line) with corresponding 68\% confidence intervals (dotted cyan lines).  
The residuals, shown in the bottom plot, have an rms of 14~G ($\chisqr=1.6$).  The six blocks of data correspond to the six observing seasons (2019, 2020, 2021, 2022, 2024 and 2025). }
\label{fig:gpb}
\end{figure}

We explored the hyperparameter domain with a Markov chain Monte Carlo (MCMC) process and retrieved posterior distributions and error bars for all parameters.  The MCMC and 
GPR modeling tools were those used in our previous studies \citep[e.g.,][]{Donati23}.  The MCMC process is a single-chain Metropolis-Hastings scheme
running over a few $10^5$ steps, including the first few $10^4$ steps as burn-in.  Convergence was checked with an autocorrelation analysis to verify that the burn-in and main
phase vastly exceeded the autocorrelation lengths of all parameters.  We computed the marginal logarithmic likelihood, $\log \mathcal{L}_M$, following \citet{Chib01}.

The achieved GPR fit is shown in Fig.~\ref{fig:gpb}, with Table~\ref{tab:gpb} listing the optimal GPR hyperparameters.  The \chisqr\ reached in the GPR fit is well 
above 1.0, witnessing the significant impact of the intrinsic variability caused by accretion on the \Bl\ data.  
The derived rotation period ($3.587\pm0.009$~d) is slightly shorter than our previous estimate and closer to that inferred from the RV curve (see Sec.~\ref{sec:rvs}).  
This difference most likely reflects differential rotation at the surface of TW~Hya, with magnetic regions closer to the pole (to which \Bl\ is mostly sensitive given the 
nearly pole-on orientation of TW~Hya) rotating more slowly 
than brightness and / or magnetic regions closer to the equator (affecting the RV curve most).  
The semi-amplitude of the \Bl\ rotational modulation was small most of the time, consistent with the orientation of TW~Hya, and largest in 2020 (see Fig.~\ref{fig:gpb}).  
The main source of variability occurred on the long term (from season to season) and also on a shorter term (within single seasons, especially in 2024 and 2025).  This faster 
evolution in the last two seasons shows up in the smaller $\theta_3$ than that inferred in our previous study.  This oscillatory behaviour of \Bl, obvious in 2024 and 2025, 
suggests that the large-scale field of TW~Hya, and thus the underlying dynamo processes amplifying the field, can include at times a QP component.  
The sign changes of \Bl\ over the first four seasons reflect local changes of the large-scale topology in an nearly pole-on viewing configuration. 

Besides, we used the median of all \texttt{APERO}-reduced spectra of TW~Hya to reassess the stellar parameters with ZeeTurbo \citep[][]{Cristofari23}
simultaneously with the small-scale magnetic field estimated from the Zeeman broadening of the spectral lines.  
The fit to the K band spectrum is shown in Fig.~\ref{fig:spc}.  For the stellar parameters, we found $\teff=3770\pm50$~K and
$\logg=4.33\pm0.10$ (assuming solar metallicity).  We also deduced that TW~Hya hosts a small-scale magnetic field of <$B$>~=~$3.2\pm0.2$~kG, with the 0, 2, 4, 6, and 8~kG 
components covering 5\%, 51\%, 32\%, 6\%, and 6\% (with error bars of a few percent).  This is consistent 
with previous results, e.g., from \citet[][$\teff=3800\pm100$~K, $\logg=4.20\pm0.10$, and <$B$>~=~$3.0\pm0.2$~kG]{Sokal18}, \citet[<$B$>~=~$3.4\pm0.2$~kG]{Lavail19}, or 
\citet[][$\teff=3783\pm107$~K, $\logg=4.35\pm0.18$, and <$B$>~=~$2.75\pm0.30$~kG]{Lopez-Valdivia23}.  The derived temperature is slightly lower than that inferred from 
optical spectra \citep[$4050\pm50$~K,][]{Donati11b,Donati24b}, presumably due to cool magnetic 
spots at the stellar surface contributing more to nIR than to optical spectra.  ZeeTurbo also derived that the veiling $r$ from accretion, reducing the depth of 
spectral lines and measured on each spectral band, was on average equal to $r=0.23\pm0.05$ throughout the median spectrum, slightly lower though still consistent with 
previous measurements in the K band \citep{Sokal18,Lavail19,Lopez-Valdivia23}.  

To investigate potential temperature changes coordinated with the modulation of \Bl, we carried out on $dT$ (found to vary from $-39$ to 44~K with a median error of 8~K) a GPR analysis 
similar to that outlined above and obtained that 
rotational modulation was too small to be reliably detected.  This mostly reflects the nearly pole-on orientation of TW~Hya, inducing only a weak modulation of $dT$, even for surface 
brightness features with a significant temperature contrast with the quiet photosphere\footnote{Line bisectors behave similarly \citep{Huelamo08}.}.

\section{ZDI modeling}
\label{sec:zdi}

We used ZDI to model the LSD Stokes $I$ and $V$ profiles of TW~Hya, under the assumption that their rotationally modulated shapes and distortions are caused by by brightness and magnetic 
surface features.  We started by correcting LSD profiles for veiling to remove, to first order, contributions from accretion funnels, flares, and the inner disk.  Owing to the clear 
evolution of the magnetic topology during the 2024 and 2025 observing seasons, each data set was divided into two subsets: 2024a (February~15 to March~02) and 2024b (March~16 to May~03) 
for 2024, and 2025a (February~06 to April~16) and 2025b (May~07 to June~10) for 2025.  We also derived relative photometric light curves at SPIRou wavelengths using the $dT$ measurements 
and converting temperature variations into brightness fluctuations through the Planck function.  These light curves were incorporated as additional constraints in the ZDI inversion 
alongside the LSD Stokes $I$ and $V$ profiles.  Compared with contemporaneous photometry from facilities such as TESS, the $dT$-based light curves have the dual advantage of being much 
less affected by accretion-related variability while matching the wavelength range of the SPIRou observations.  In all seasons, the resulting light curves are nearly flat, consistent 
with the absence of detectable rotational modulation in $dT$ (see Sec.~\ref{sec:mag}), with full amplitudes at SPIRou wavelengths of only a few tenths of a percent.  We emphasize that 
simultaneously fitting both the LSD Stokes $I$ and $V$ profiles allows the ZDI modeling to account for Zeeman broadening, thereby avoiding a significant underestimate of the reconstructed 
magnetic field strength \citep[e.g.,][]{Donati25b}.  Finally, to ensure internal consistency, we reapplied ZDI to the data from the previous observing seasons using the updated rotation 
period, the revised modeling parameters (see below), and the inclusion of the $dT$-derived photometric constraints, relative to our earlier analysis \citep{Donati24b}.

In practice, ZDI starts from empty magnetic and brightness maps.  Assuming solid-body rotation, the code iteratively updates both surface distributions, exploring 
the parameter space with a conjugate-gradient algorithm.  At each iteration, synthetic Stokes profiles computed from the current maps are compared with the observations until the target 
level of agreement, quantified by a prescribed \chisqr, is achieved.  The stellar surface is discretized into 5000 grid cells, each assigned a local surface brightness relative to the 
quiet photosphere.  The magnetic field is represented through a spherical harmonics (SH) expansion following the formalism of \citet{Donati06b} in its revised implementation 
\citep{Lehmann22,Finociety22}.  The poloidal and toroidal components are described by three sets of complex SH coefficients: $\alpha_{\ell,m}$ and $\beta_{\ell,m}$ for the poloidal field, 
and $\gamma_{\ell,m}$ for the toroidal field, where $\ell$ and $m$ denote the degree and order of the corresponding SH mode.  For TW~Hya, the expansion was truncated at $\ell=5$, 
consistent with its low projected rotational velocity (\vsini~=~3~\kms), which limits the spatial resolution achievable by ZDI.  Since the number of model parameters exceeds the number 
of independent observational constraints, the inversion is intrinsically ill-posed and must be regularized.  ZDI selects, among all solutions fitting the data at the 
specified \chisqr, the one with the minimum information content (or equivalently, maximum entropy), following the prescription of \citet{Skilling84}.

\begin{figure*}[ht!]
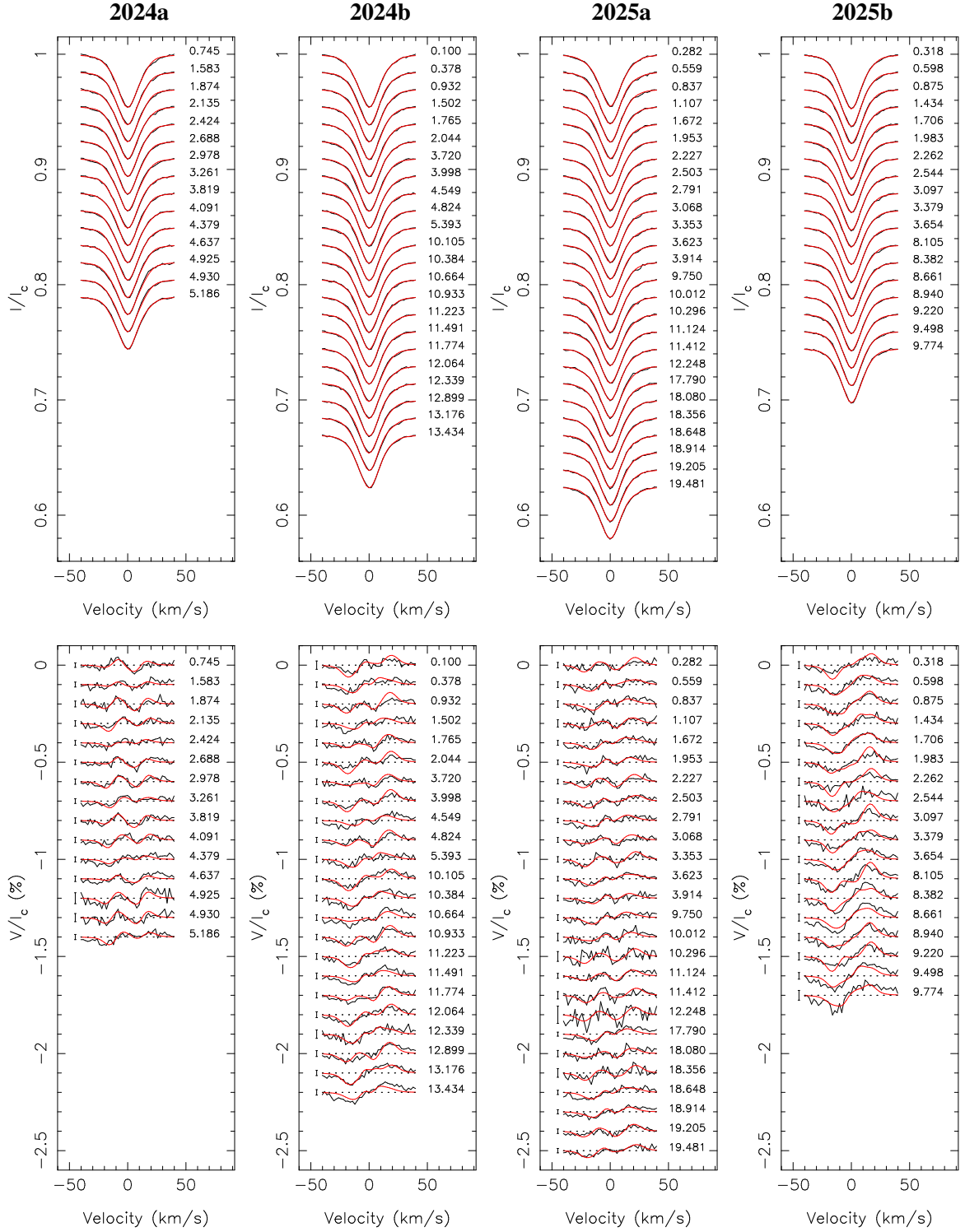
 
\flushleft{\large\bf \hspace{3.2cm}2024a\hspace{2.9cm}2024b\hspace{2.9cm}2025a\hspace{2.9cm}2025b\vspace{-4mm}}
\center{\includegraphics[scale=0.28,angle=-90]{fig2/twhya2-fiti24a.ps}\hspace{2mm}\includegraphics[scale=0.28,angle=-90]{fig2/twhya2-fiti24b.ps}\hspace{2mm}\includegraphics[scale=0.28,angle=-90]{fig2/twhya2-fiti25a.ps}\hspace{2mm}\includegraphics[scale=0.28,angle=-90]{fig2/twhya2-fiti25b.ps}\vspace{1mm}}   
\center{\includegraphics[scale=0.28,angle=-90]{fig2/twhya2-fitv24a.ps}\hspace{2mm}\includegraphics[scale=0.28,angle=-90]{fig2/twhya2-fitv24b.ps}\hspace{2mm}\includegraphics[scale=0.28,angle=-90]{fig2/twhya2-fitv25a.ps}\hspace{2mm}\includegraphics[scale=0.28,angle=-90]{fig2/twhya2-fitv25b.ps}}   
\caption[]{Observed (thick black line) and modeled (thin red line) LSD Stokes $I$ (top row) and $V$ (bottom row) profiles of TW~Hya in epochs 2024a, 2024b, 2025a and 2025b (from left to right).
Rotation cycles (counting from 492, 501, 592, and 617 for 2024a, 2024b, 2025a, and 2025b, respectively, see Table~\ref{tab:log}) are indicated to the right of the LSD profiles, and $\pm$1$\sigma$ 
error bars are shown to the left of the Stokes $V$ profiles.  The discrepant fits to some Stokes $V$ profiles, especially in 2025, reflects the significant intrinsic evolution of the magnetic 
topology within a single data set (see, e.g., the differing Stokes $V$ profiles at the nearly equal rotation phases 3.379 and 8.382 in 2025b) that a static magnetic model cannot reproduce.  }
\label{fig:fit}
\end{figure*}

\begin{figure*}[ht!]
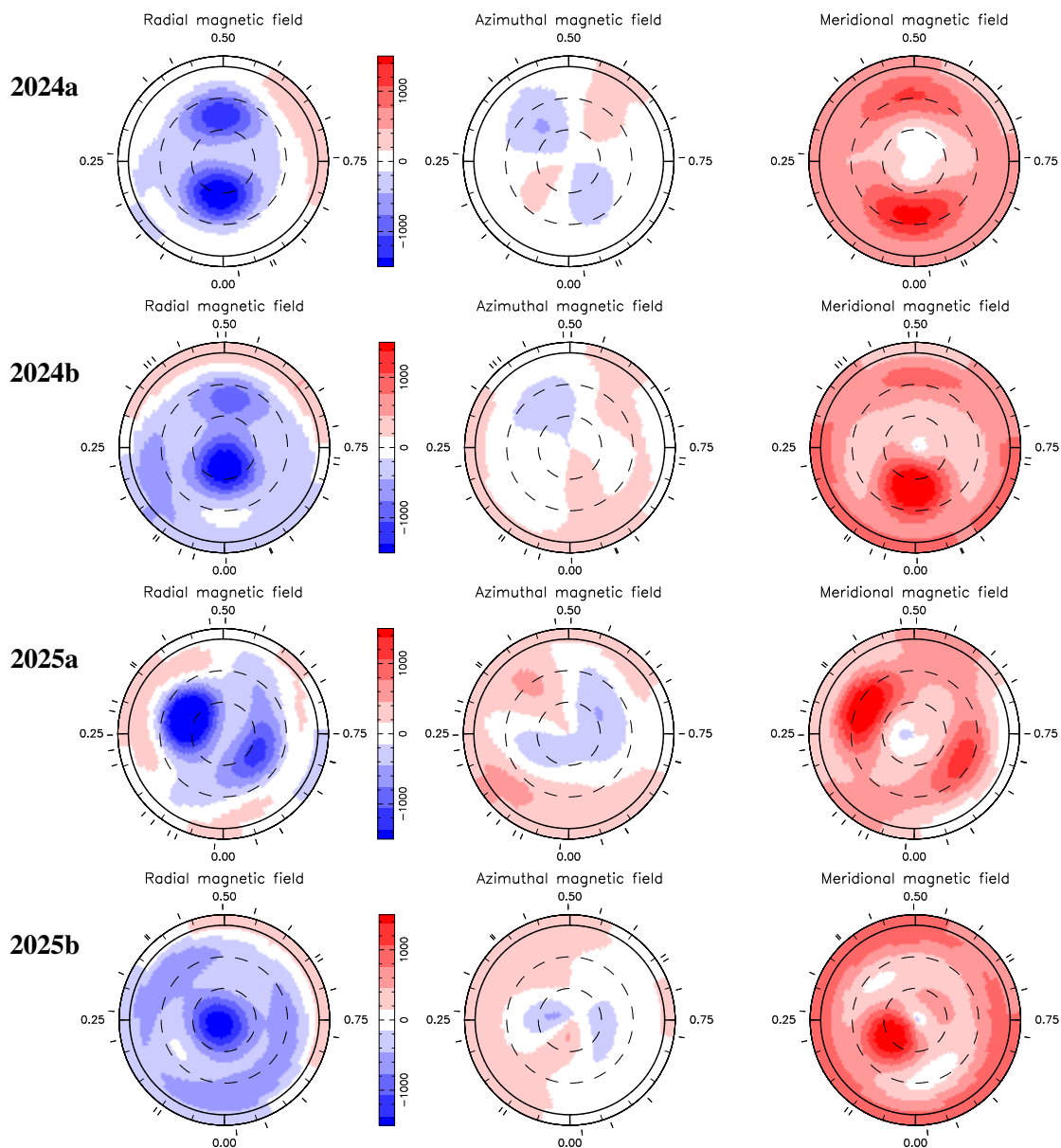
 
\centerline{\large\bf 2024a\raisebox{0.30\totalheight}{\includegraphics[scale=0.38,angle=-90]{fig2/twhya2-map24a.ps}}\vspace{1mm}}
\centerline{\large\bf 2024b\raisebox{0.30\totalheight}{\includegraphics[scale=0.38,angle=-90]{fig2/twhya2-map24b.ps}}\vspace{1mm}}
\centerline{\large\bf 2025a\raisebox{0.30\totalheight}{\includegraphics[scale=0.38,angle=-90]{fig2/twhya2-map25a.ps}}\vspace{1mm}}
\centerline{\large\bf 2025b\raisebox{0.30\totalheight}{\includegraphics[scale=0.38,angle=-90]{fig2/twhya2-map25b.ps}}} 
\caption[]{Maps of the large-scale field at the surface of TW~Hya reconstructed with ZDI for epochs 2024a, 2024b, 2025a, and 2025b (top to bottom rows, respectively) 
from the LSD Stokes $I$ and $V$ profiles of Fig.~\ref{fig:fit}, showing the radial, azimuthal, and meridional 
components in spherical coordinates (left, middle and right columns, units in G).  Maps are displayed in a flattened polar projection 
down to latitude $-10$\degr, with the north pole at the center, the equator shown as a bold line, and outer ticks marking phases of 
observations.   Negative radial, azimuthal, and meridional fields point inward, clockwise, and equatorward, respectively. }
\label{fig:map}
\end{figure*} 

Local synthetic Stokes $I$ and $V$ profiles are computed for each surface element using the Unno-Rachkovsky solution to the polarized radiative transfer equation in a plane-parallel Milne-Eddington 
atmosphere \citep{Landi04}, adopting a linear limb-darkening law with a coefficient of 0.3.  At each observed rotational phase, the contributions from all visible surface elements are 
integrated to produce the synthetic disk-integrated profiles, whose mean wavelength (1750~nm) and effective Land\'e factor (1.2) match those of the observed LSD profiles.  Throughout this study, we 
adopted a local Doppler width of $\vD=3.0$~\kms, representative of TW~Hya and similar stars. As in previous ZDI studies, we introduced two filling factors, assumed uniform over the stellar surface: 
$f_V$ for the large-scale magnetic field sampled by the Stokes $V$ profiles, and $f_I$ for the small-scale magnetic field responsible for the Zeeman broadening of the Stokes $I$ profiles.  Accordingly, 
a surface element reconstructed with a magnetic field strength $B$ is assumed to contain a field of strength $B/f_V$ occupying a fraction $f_V$ (or $f_I$) of its area when modelling the Stokes $V$ (or 
$I$) profiles.  This prescription allows Zeeman broadening to be consistently included in the modelling of both Stokes $I$ and $V$ profiles, while ensuring that the ratio of the large-scale to the 
small-scale magnetic flux, given by $f_V/f_I$, remains compatible with values inferred for active low-mass stars \citep[e.g.,][]{Morin10,Kochukhov21}.  We adopted $f_I=0.9$, in agreement with the 
ZeeTurbo analysis (see Sec.~\ref{sec:mag}), and $f_V=0.3$, slightly lower than in our previous study but more consistent with values derived for comparable young stars \citep[e.g.,][]{Donati25c}. 
Synthetic light curves are computed analogously by summing the contributions of all visible surface elements at each rotational phase, with each contribution determined by the local 
brightness and limb angle.

\begin{table}[t!]
\caption{Large-scale and small-scale fields measured from the magnetic topologies derived with ZDI in all seasons / subsets}
\centering
\resizebox{\linewidth}{!}{
\begin{tabular}{cccccccc}
\hline
         & \multicolumn{6}{c}{Stokes $I$ \& $V$ analysis ($f_I=0.9$, $f_V=0.3$, $\vD=3.0$~\kms)}    \\
\hline
Season   & <$B_V$> & <$B_I$> & <$B_s$> & \Bd  & tilt / phase & \Bo  & pol/axi   \\
         &  (kG)   & (kG)    & (kG)    & (kG) & (\degr / )   & (kG) & (\%)      \\
\hline
2019     & 0.83   & 2.5   & 2.7   & $-$0.67 &  11 / 0.39 & 0.30 & 95 / 87 \\
2020     & 0.99   & 3.0   & 2.9   & $-$1.07 &  29 / 0.96 & 0.49 & 92 / 75 \\
2021     & 0.89   & 2.7   & 2.8   & $-$0.86 &  21 / 0.60 & 0.44 & 93 / 81 \\
2022     & 0.83   & 2.5   & 2.7   & $-$0.72 &  19 / 0.26 & 0.35 & 95 / 85 \\
2024a    & 0.83   & 2.5   & 2.6   & $-$0.78 &  16 / 0.20 & 0.38 & 97 / 83 \\
2024b    & 0.90   & 2.7   & 2.8   & $-$0.89 &  15 / 0.06 & 0.23 & 91 / 85 \\
2025a    & 0.89   & 2.7   & 2.8   & $-$0.81 &   7 / 0.30 & 0.44 & 91 / 79 \\
2025b    & 0.87   & 2.6   & 2.6   & $-$0.88 &  16 / 0.10 & 0.18 & 90 / 92 \\
\hline
\end{tabular}}
\tablefoot{Columns 2 and 3 list the quadratically averaged large-scale field (<$B_V$>) and small-scale field (<$B_I$>) over the stellar surface.
Column~4 gives the time-averaged small-scale field integrated over the visible hemisphere <$B_s$> (to be compared with the average <$B$> measured with ZeeTurbo, 
see Sec.~\ref{sec:mag}).  Columns~5 to 8 list the polar strength of the dipole component \Bd, the tilt of the dipole component to the rotation axis and the phase toward
which it is tilted, the polar strength of the octupole component, and the amount of magnetic energy reconstructed in the poloidal component of the field and in the axisymmetric 
modes of this component.  Typical error bars on field values, fractional amounts of magnetic energy, and dipole tilts are about 0.1~kG, 5\%, and 5\degr. }
\label{tab:mag}
\end{table}

Figure~\ref{fig:fit} presents the ZDI fits to the LSD Stokes $I$ and $V$ profiles for the four most recent data subsets (2024a to 2025b), while Fig.~\ref{fig:map} shows the corresponding 
reconstructed magnetic maps, obtained assuming $\vsini=3.0$~\kms\ \citep[as in][]{Donati24b}. For comparison, the maps reconstructed from the 2019 to 2022 data are displayed in Fig.~\ref{fig:map2}. 
The main magnetic properties of all reconstructed maps are summarized in Table~\ref{tab:mag}.  In every observing season, the recovered magnetic topology is predominantly poloidal and axisymmetric, 
with the dipole component ranging from 0.67 to 1.07~kG and accounting for approximately 80\% of the reconstructed poloidal magnetic energy.  The octupolar component exhibits a more complex 
morphology and has an average strength of only about 40\% that of the dipole.  Despite temporal variations, the reconstructed maps display several recurring features.  In particular, the visible 
rotational pole generally hosts a split region of negative radial field.  The four most recent subsets also reveal significant evolution of the large-scale magnetic topology within a single 
observing season, in agreement with the variations seen in the $\Bl$ curve (see Sec.~\ref{sec:mag}).  This evolution is especially pronounced in 2025, when the two negative-polarity patches near 
the visible pole merged into a single structure within only a few weeks.  Such rapid evolution likely explains the slightly poorer fits to the Stokes $V$ LSD profiles obtained for the 2025 subsets 
(Fig.~\ref{fig:fit}), reflecting departures from the assumption of a static magnetic topology over each subset.  The mean small-scale magnetic field reconstructed by ZDI, <$B_s$>, ranges from 2.6 
to 2.9~kG (Table~\ref{tab:mag}).  These values are slightly lower, but still consistent within the uncertainties, with the average field strength inferred from ZeeTurbo, <$B$>$=3.2\pm0.2$~kG 
(Sec.~\ref{sec:mag}).  Moreover, <$B_s$> remains nearly constant throughout the observing campaign, mirroring the stability of $dT$, which has been shown to provide a reliable proxy for <$B$> 
\citep{Artigau24,Cristofari25}.

The brightness maps reconstructed simultaneously with the magnetic maps (not shown) contain essentially no bright or dark features with contrasts exceeding 10\% of the quiet photospheric flux, 
in agreement with the nearly flat photometric light curves inferred from the $dT$ measurements.  We nevertheless expect that TW~Hya hosts cool spots on its visible hemisphere, particularly near 
the magnetic pole, as suggested for instance by the difference between the effective temperatures derived from the optical and near-infrared spectra (see Sec.~\ref{sec:obs}). However, owing to 
the nearly pole-on inclination of the system and its low projected rotational velocity, these spots produce only weak rotational signatures in the Stokes $I$ profiles and $dT$ measurements. 
As a result, they have little influence on the ZDI inversion and need not be explicitly reconstructed.

\section{RV modeling}
\label{sec:rvs}

As in \citet{Donati24b}, we examined RVs derived by LBL, whose error bars based on photon noise range from 1.8 to 6.0~\ms\ (median 2.2~\ms).  We found that these RVs 
exhibited a clear trend with BERV, getting $\simeq$50~\ms\ smaller for BERV values around 15~\kms\ (i.e., close to the stellar RV of 12.4~\kms) as a result of a 
spectral contamination by telluric residuals (in particular OH emission sky lines).  To remove this trend 
without rejecting data points, we carried out a GPR fit to the RV vs.\ BERV data using a purely exponential covariance function\footnote{In this case, we achieved this by arbitrarily 
fixing $\theta_2=1000$~d and $\theta_4=1$ in Eq.~\ref{eq:covar}, as in, e.g., \citet{Donati23b}.}, then subtracted the derived GP from the RV data to remove the trend with BERV.  
We ended up with the corrected LBL RVs listed in Table~\ref{tab:log}.  A very similar result was obtained when applying the \texttt{WAPITI} filtering technique aimed at correcting 
residual telluric pollution in SPIRou RVs \citep{Ould-Elhkim23}.  

\begin{figure}[ht!]
\centerline{\includegraphics[scale=0.38,angle=-90]{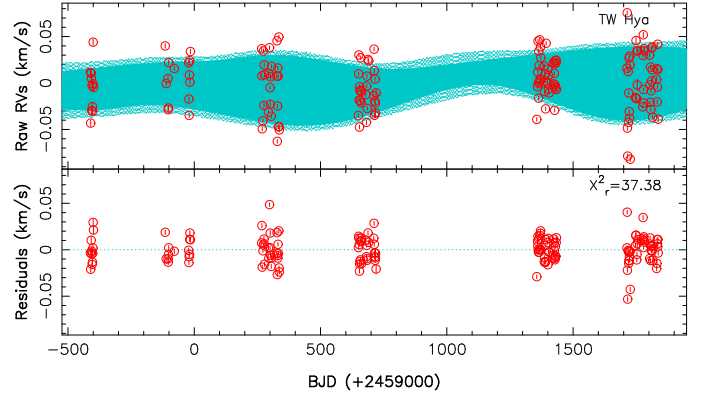}}
\caption[]{BERV-corrected LBL RVs of TW~Hya (red dots) over all observing seasons.  The top and bottom panels respectively show the MCMC GPR fit to the data (modeling the activity, cyan curve) 
and the residuals (rms 13.5~\ms). } 
\label{fig:rv}
\end{figure}

\begin{figure}[ht!]
\centerline{\includegraphics[scale=0.38,angle=-90]{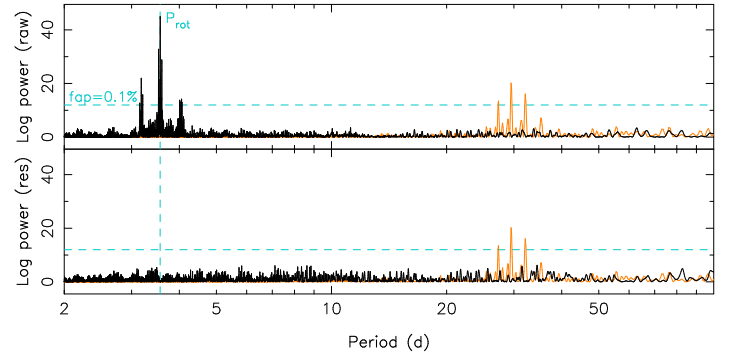}}
\caption[]{Periodogram of the raw (top) and residual (bottom) BERV-corrected LBL RV data following a MCMC GPR fit to the data (modeling the activity, see Fig.~\ref{fig:rv}).  
The dashed vertical cyan line traces the stellar rotation period and the dashed horizontal lines indicate a 0.1\% FAP level in the periodogram of the RV data.
The orange curve depicts the periodogram of the window function.  } 
\label{fig:per}
\end{figure}

In a second step, we ran a MCMC GPR fit to the corrected LBL RVs with a QP kernel, to model the activity jitter induced by the rotation of the star, leaving this time all 
five GPR hyperparameters free to vary.  The results are shown in Fig.~\ref{fig:rv}, with the derived hyperparameters listed in Table~\ref{tab:gprv}.  The activity jitter 
is clearly detected with an average semi-amplitude of $32\pm11$~\ms, well above the residuals (rms 13.5~\ms).  We also find significant excess RV noise ($14.9\pm0.9$~\ms), 
presumably resulting from stochastic accretion and activity (e.g., flares).  Figure~\ref{fig:per} presents a periodogram of the raw and residual RVs, 
with an obvious peak at the rotation period in the raw RVs and no peak getting close to the 0.1\% false-alarm probability (FAP) threshold in the residual RVs that could suggest the 
presence of a close-in planet with an orbital period in the range 5--100~d.  In particular, we were not able to confirm the existence of the candidate planet at an orbital period 
of $\simeq$8~d suggested in our previous study, whose signal was presumably linked to the spurious RV trend with BERV that we filtered in the present study.  

\begin{figure*}[ht!]
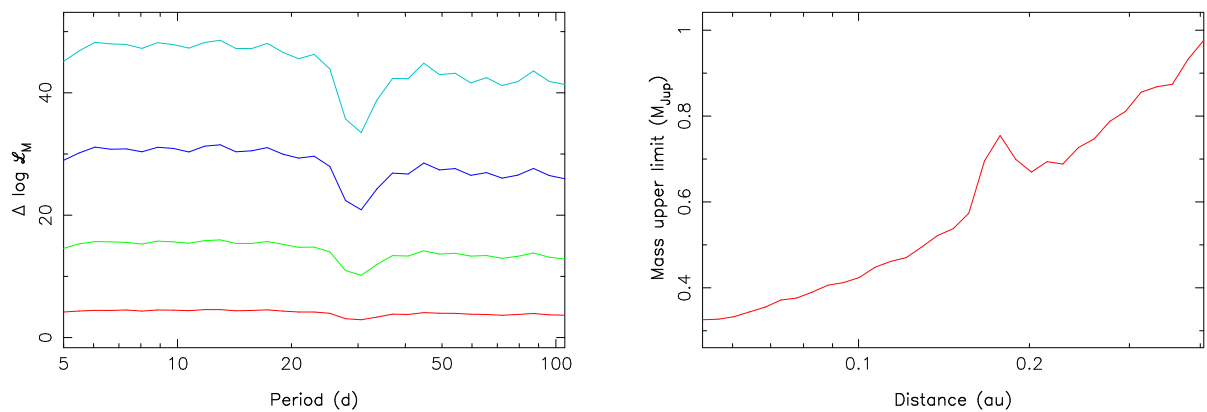

\centerline{\includegraphics[scale=0.32,angle=-90]{fig2/twhya2-pldet.ps}\hspace{10mm}\includegraphics[scale=0.32,angle=-90]{fig2/twhya2-msup.ps}} 
\caption[]{Detectability of close-in planets around TW~Hya.  Left panel: Logarithmic Bayes factor $\Delta \log \mathcal{L}_M$ as a function of orbital period for planet RV signatures of 
semi-amplitudes 5 (red), 10 (green), 15 (blue), and 20~\ms\ (cyan) once averaged over all phases.  Note the drop in $\Delta \log \mathcal{L}_M$ at orbital periods close to the 
synodic period of the Moon. Right panel: mass upper limit (corresponding to $\Delta \log \mathcal{L}_M=10$) as a function of orbital distance, of a close-in planet in a circular orbit 
inclined at 10\degr\ to the line of sight.  }  
\label{fig:det}
\end{figure*}

Finally, we ran injection recovery tests to investigate the semi-amplitude threshold $K$ above which detections can be considered as reliable, assuming a close-in planet in a circular 
orbit with orbital periods ranging from 5 to 100~d (see Fig.~\ref{fig:det}, left panel). 
We found that, for periods in the range 5$-$20~d, $K\simeq7.5$~\ms\ is required on average to ensure a reliable detection with $\Delta \log \mathcal{L}_M=10$, which corresponds to minimum 
masses of the putative planet of 0.054$-$0.086~\mjup\ depending on the period, i.e., planet masses of 0.31$-$0.50~\mjup\ assuming an orbital axis inclination to the line of sight equal 
to that of the stellar rotation axis ($i=10$\degr).  Beyond the drop in sensitivity for orbital periods close to the synodic period of the Moon (for which $K\simeq9.9$~\ms, corresponding 
to a minimum mass of 0.13~\mjup, is required for a safe detection at an orbital period of 30~d), the planet detectability weakens slightly compared to that at short periods.  
$K\simeq8.2$~\ms\ is needed to ensure a secure detection for periods in the range 40$-$100~d, which corresponds to minimum planet masses of 0.12$-$0.16~\mjup (i.e., planet masses 
of 0.68$-$0.93~\mjup).  
Altogether, our results imply that the upper limit for the mass of a potential close-in exoplanet in the inner disk of TW~Hya ranges from 0.33 to 0.98~\mjup\ for distances 0.053 to 0.41~au, 
with a local peak at 0.75~\mjup\ at an orbital distance of 0.18~au (see Fig.~\ref{fig:det}, right panel).  
Beyond orbital periods of 100~d, the planet detectability rapidly degrades as a result of the poor temporal sampling, with upper limits on the planet mass of about 3~\mjup\ at 1~au and 
15~\mjup\ at 2~au.

\begin{table}[t!]
\caption{Same as Table~\ref{tab:gpb} for the BERV-corrected LBL RVs of TW~Hya}
\centering
\resizebox{\linewidth}{!}{
\begin{tabular}{cccc}
\hline
Parameter   & Symbol & Value & Prior   \\
\hline
GP amplitude (\kms)   & $\theta_1$  & $0.032\pm0.011$   & mod Jeffreys ($\sigma_{\rm RV}$) \\
Rec.\ period (d)      & $\theta_2$  & $3.5635\pm0.0014$ & Gaussian (3.6, 0.5) \\
Evol.\ timescale (d)  & $\theta_3$  & $372\pm88$        & log Gaussian ($\log$ 350, $\log$ 2.0) \\ 
Smoothing             & $\theta_4$  & $0.91\pm0.32$     & Uniform  (0, 3)    \\
White noise (\kms)    & $\theta_5$  & $0.0149\pm0.0009$ & mod Jeffreys ($\sigma_{\rm RV}$) \\
Rms (\kms)            &             & 0.0135            & \\ 
$\chisqr$             &             & 37                & \\
\hline
\end{tabular}}
\label{tab:gprv}
\end{table}

\section{Accretion and activity}
\label{sec:act}

Even though TW~Hya is not a strong accretor, its spectrum is subject to veiling, especially at optical wavelengths \citep{Herczeg23,Ji26} but also in the nIR \citep{Sousa23}.  
We estimated the veiling by comparing the equivalent widths (EW) of the LSD Stokes $I$ profiles of TW~Hya with those of the weak-line T~Tauri star V819~Tau, whose spectral 
type is similar \citep[see][]{Sousa23}.  Using atomic lines concentrating mostly in the $JH$ band, we derived the corresponding veiling $r_{JH}$;  we obtained a 
second estimate in the $K$ band, $r_K$, from the CO bandhead lines (redward of 2.28~\mic).  The values we inferred, listed in Table~\ref{tab:log}, ranged from 0.1 to 0.5 for $r_{JH}$ 
(median 0.24), and from 0.0 to 0.7 for $r_K$ (median 0.20).  Overall, the veiling was small, even in the K band where the largest values potentially reflect incomplete removal 
of the thermal background polluting the K band.  The Pearson correlation coefficient between $r_K$ and $r_{JH}$ was equal to $R\simeq0.60$ (with $r_K$ being on average 
equal to $r_{JH}$), implying a fair correlation given the measurement error (of 0.05$-$0.10).  Neither $r_{JH}$ nor $r_K$ showed rotational modulation, each exhibiting multiple 
weak peaks in the periodogram.  We note that the average veiling in 2024 was the weakest of all seasons (0.17 for both $r_{JH}$ and $r_K$), suggesting that 2024 
was a period of particularly low accretion rates for TW~Hya.  

\begin{figure*}[ht!]
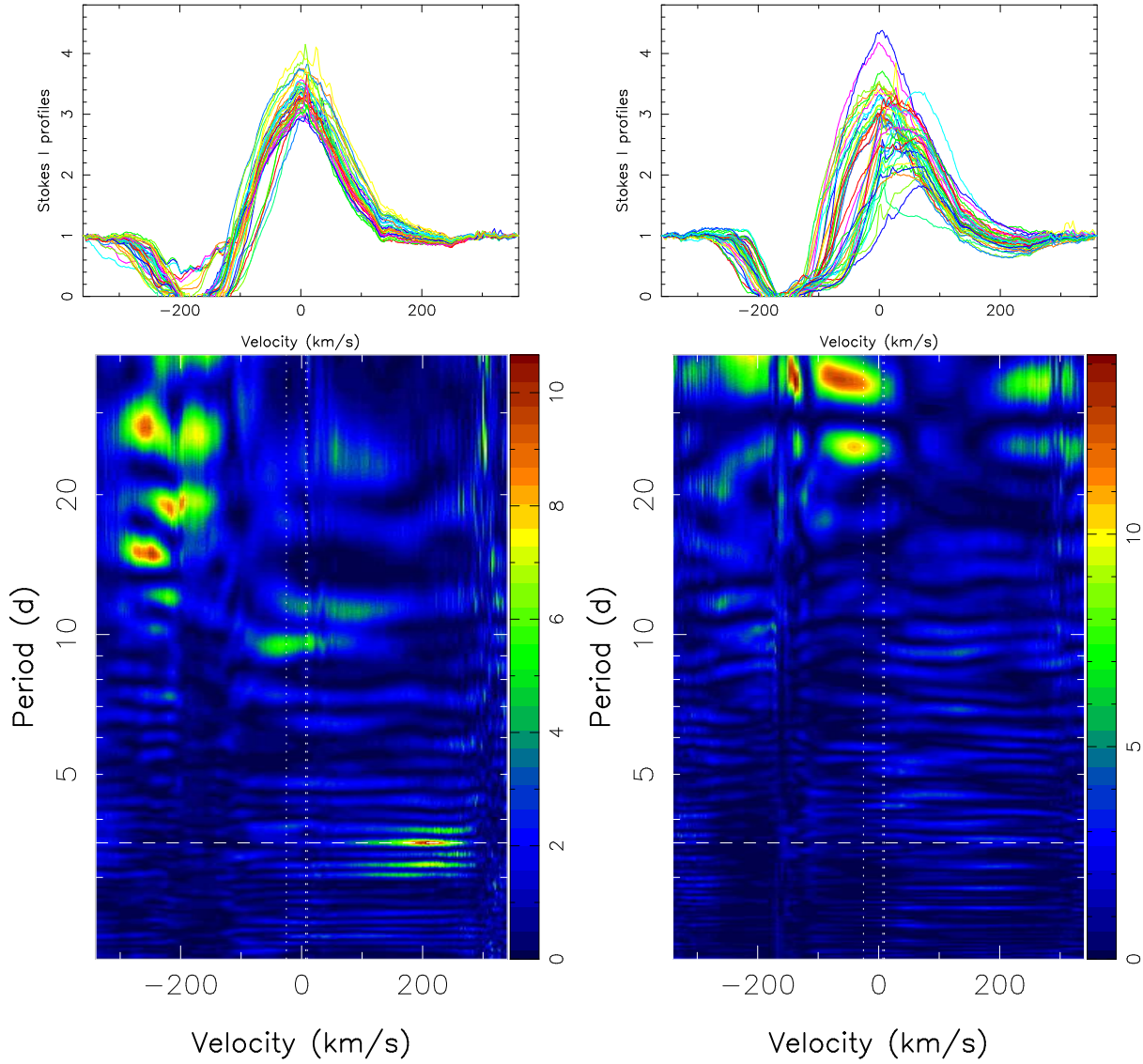

\resizebox{0.93\textwidth}{!}{
\centerline{\hspace{-2mm}\includegraphics[scale=0.32,angle=-90]{fig2/twhya2-hei24.ps}\hspace{14.5mm}\includegraphics[scale=0.32,angle=-90]{fig2/twhya2-hei25.ps}\vspace{2mm}}}
\resizebox{0.93\textwidth}{!}{
\centerline{\includegraphics[scale=0.55,angle=-90]{fig2/twhya2-hei24per-2.ps}\hspace{3mm}\includegraphics[scale=0.55,angle=-90]{fig2/twhya2-hei25per-2.ps}}}
\caption[]{
Stacked Stokes $I$ profiles (top plots) and 2D periodograms (bottom plots) of the 1083.3~nm \hei\ IRT in the stellar rest frame for seasons 2024 (left panels) and 2025 (right panels).  
The dashed horizontal line traces \Prot\ and the vertical dotted lines depict the velocities of the three components of the \hei\ triplet.  The color-scale traces the logarithmic 
power in the periodogram.  Only the main peaks (yellow to red and extending over at least several velocity bins) are likely to be significant. }
\label{fig:hei}
\end{figure*}

We examined the 1083~nm \hei\ triplet and constructed the 2D periodograms of its profile for seasons 2024 and 2025 (see Fig.~\ref{fig:hei}).  In 2025, the observing season was 
long enough to allow us to compute meaningful periodograms for both 2025a and 2025b (see Fig.~\ref{fig:hei2}).  As in our previous study, the 
line profile featured strong blueshifted absorption probing the stellar wind (extending down to $-300$~\kms), as well as redshifted emission associated with accretion processes 
(extending to 250~\kms).  It also featured weak though nonetheless clear redshifted absorption centred at 200~\kms\ and tracing accretion funnels regularly crossing the line 
of sight.  Rotational modulation was detected very clearly in the redshifted absorption in 2024 but not in 2025 (or only very weakly in the second half of the season, see 
Fig.~\ref{fig:hei2}), although redshifted absorption was stronger in 2025 than in 2024.  From the temporal evolution of the 2D periodogram in 2025 (see Fig.~\ref{fig:hei2}), we 
can confirm that the accretion pattern between the disk and 
the star was unstable, likely as a result of the contemporaneous evolution of the large-scale field topology (see Sec.~\ref{sec:zdi}) and / or of the clumpy nature of the inner 
disk \citep{Ji26}.  Longer periods were also present in the blue wing of the blueshifted component probing the stellar wind, e.g., at $\simeq$15~d in 2024, and at $\simeq$10~d in 
the central emission (also in 2024).  Most of the other periodogram peaks at periods of 25$-$35~d are likely attributable to sampling aliases (see orange curve in Fig.~\ref{fig:per}).  

\begin{figure*}[ht!]
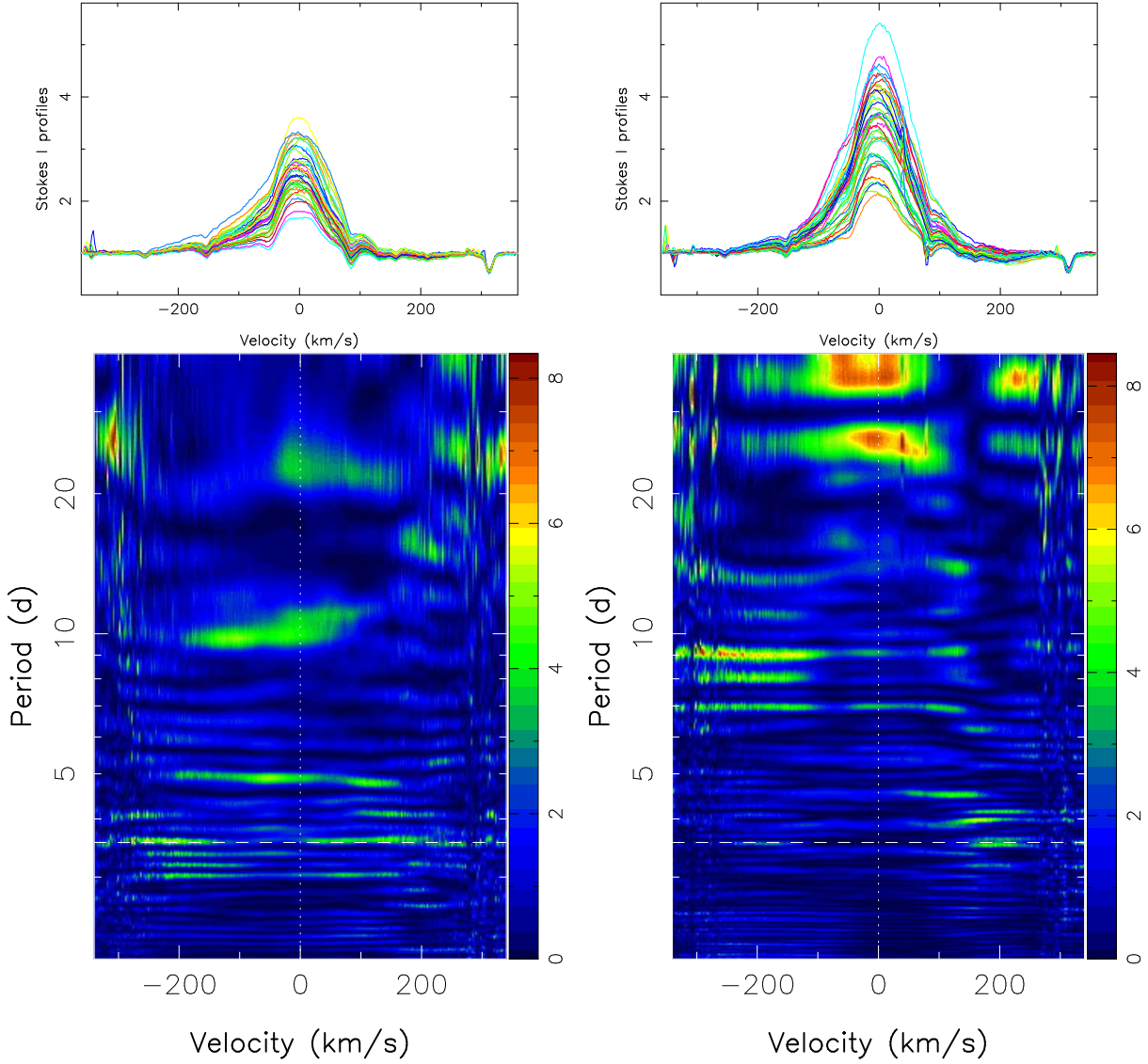

\resizebox{0.93\textwidth}{!}{
\centerline{\hspace{-2mm}\includegraphics[scale=0.32,angle=-90]{fig2/twhya2-pab24.ps}\hspace{14.5mm}\includegraphics[scale=0.32,angle=-90]{fig2/twhya2-pab25.ps}\vspace{2mm}}}
\resizebox{0.93\textwidth}{!}{
\centerline{\includegraphics[scale=0.55,angle=-90]{fig2/twhya2-pab24per-2.ps}\hspace{3mm}\includegraphics[scale=0.55,angle=-90]{fig2/twhya2-pab25per-2.ps}}}
\caption[]{Same as Fig.~\ref{fig:hei} for the 1282~nm \pab\ line (left: 2024, right: 2025)}
\label{fig:pab}
\end{figure*}

We similarly looked at the 1282~nm \pab\ and 2166~nm \brg\ lines, whose profiles and 2D periodograms for seasons 2024 and 2025 are shown in Figs.~\ref{fig:pab} and \ref{fig:brg}. 
Both lines exhibited similar patterns, \brg\ being weaker and noisier than \pab.  In particular, both \pab\ and \brg\ were significantly stronger in 2025 than in 2024, in 
agreement with season-averaged veiling measurements.  We also detected redshifted absorption in \pab, stronger in 2025 than in 2024 (as for the \hei\ triplet), with a weak level 
of rotational modulation at both epochs.  Again, the 2D periodogram of \pab\ evolved significantly within the 2025 season, with rotational modulation of the redshifted absorption 
being clear in 2025b but not in 2025a (with power detected at a period or 3.9~d, see Fig.~\ref{fig:pab2}).  Additional periods of 9$-$10~d were also detected 
in the blueshifted half of \pab\ (and \brg\ to a lower extent).  

By summing the 150$-$250~\kms\ spectral bins of the \hei\ and \pab\ lines, we obtained fluxes in the redshifted absorption region of the lines.  As readily visible on the 2D 
periodograms, these fluxes were rotationally modulated, in 2024 for the \hei\ line (with a period of $3.55\pm0.02$~d) and in 2025b for \pab\ (with a period of 
$3.56\pm0.01$~d).  The maximum \hei\ redshifted absorption occurred at phase $0.5\pm0.1$ in 2024, coinciding with the location of the negative radial field spot that weakened 
in this season (see Fig.~\ref{fig:map}).   For \pab\ in 2025b, the maximum redshifted absorption also took place at phase $0.5\pm0.1$, but it is not obvious 
to relate it to a specific magnetic feature given that the large-scale magnetic topology was almost fully axisymmetric.  That the period of the \pab\ redshifted absorption was 
significantly different from the rotation period in 2025a (3.9~d, see Fig.~\ref{fig:pab2}) is further evidence that the accretion pattern was 
progressively readjusting to the undergoing evolution of the large-scale field (see Fig.~\ref{fig:map}).  

We computed the emission EWs for \pab\ and \brg\ (see Table~\ref{tab:log}) and found that they correlated well with one another ($R=0.97$) but not so much with $r_{JH}$ 
($R\simeq0.4$) and even less with $r_K$ ($R\simeq0.2$).  We found that \pab\ EWs are stronger than \brg\ ones by an average factor of 4.2.  We then used the scaling relations of 
\citet{Fiorellino25} and our stellar parameters to translate the veiling-corrected EWs of \pab\ and \brg\ into accretion fluxes and the logarithmic mass-accretion rates $\log \Mdot$ 
for each visit.  The time-averaged values we obtained are $\log \Mdot = -8.25\pm0.28$~dex (with \Mdot\ in \mspy) for \pab\ and $-8.42\pm0.28$~dex for \brg, the error bar denoting the 
dispersion over all visits.  Taking the weighted average of both estimates yielded $\log \Mdot = -8.33\pm0.20$~dex, marginally larger than the values derived in our previous study 
\citep[as a result of using the updated scaling relations of][]{Fiorellino25} and than previous estimates \citep[e.g.,][]{Herczeg23}.  We note in particular that the average $\log \Mdot$ 
varied between seasons, $-8.14\pm0.12$ in 2019, $-7.95\pm0.10$ in 2020, $-8.39\pm0.16$ in 2021, $-8.18\pm0.08$ in 2022, $-8.59\pm0.16$ in 2024 and $-8.30\pm0.17$ in 2025.  
Season 2024 clearly appeared as the one during which TW~Hya exhibited the weakest accretion rate since 2019.  The change in $\log \Mdot$ within 2024 is insignificant (a decrease of 
0.04~dex between 2024a and 2024b) whereas that within 2025 is larger (increase of 0.12~dex between 2025a and 2025b) yet typical of the dispersion within one season.  
As for veiling, we found no rotational modulation of \Mdot\ over the whole data set, consistent with the nearly pole-on geometry and the intrinsic variability of accretion from the 
inner edge of the disk.  Even on the short periods when redshifted absorption of \pab\ exhibited rotational modulation (e.g., 2025b), \Mdot\ was not rotationally modulated \citep[in 
contrast with previous reports,][]{Pittman25}.  

From the time-averaged mass-accretion rate, the intensity of the reconstructed dipole component of the large-scale field ($0.83\pm0.11$~kG; see Table~\ref{tab:mag}) and the stellar 
parameters assumed for TW~Hya ($\mstar=0.8$~\msun\ and $\rstar=1.16$~\rsun), we found, following \citet{Bessolaz08}, that the magnetospheric gap of TW~Hya extended to a distance of 
$\rmag = 3.2\pm0.5$~\rstar\ or $0.017\pm0.003$~au (the error bar corresponding to variations of \Mdot\ and \Bd).  
This implies that $\rmag/\rcor = 0.40\pm0.06$ where \rcor\ denotes the corotation radius (located at 7.9~\rstar, i.e., 0.043~au) at which the Keplerian angular velocity equals the 
rotation rate at the stellar surface\footnote{Error bars on \rstar\ and \mstar\ further contribute to the uncertainty on \rmag/\rcor\ at a level of 0.08.}.  
The seasonal changes of $\rmag/\rcor$ ranged from 0.31 in 2019 (weakest \Bd, possibly reflecting the sparser data set in this season) 
up to 0.47 in 2024 (weakest \Mdot), with intermediate values of 0.36 (2020), 0.42 (2021), 0.33 (2022) and 0.39 (2025).  The latest 3D MHD simulations of \citet{Romanova25b} dedicated 
to the specific case of TW~Hya suggest a slightly smaller (though still consistent) average magnetospheric radius $\rmag/\rcor=0.33\pm0.06$ for the \Bd\ and \Mdot\ values quoted above 
\citep[implying $\mu\simeq0.5$, see Table~3 in][]{Romanova25b}.  This confirms that TW~Hya was in a regime of unstable accretion, with one or two tongues of material from the inner disk 
penetrating the magnetosphere \citep{Blinova16}.  This might explain in particular the evolving accretion patterns observed in \pab\ and \brg, and in particular the rapid fluctuations 
in the redshifted absorption component and its changing periodicity (see Fig.~\ref{fig:pab2}).  
Our estimate of $\rmag/\rcor$ is smaller than that derived from interferometric measurements from 2019 data \citep[$\rmag/\rcor\simeq0.5$;][]{Gravity20}, with both confirming 
that TW~Hya accretes material from the inner disk in an unstable fashion \citep{Blinova16}.  

We also looked at the TESS light curve from 2025 (sector 90) as obtained by {\tt TESSExtractor} \citep{Serna21}, and collected contemporaneously with our SPIRou data, and derived 
the stacked periodogram (see Fig.~\ref{fig:tes}) which showed 
no hint of the rotation period, but rather a signal at a period of 6.2~d that we did not detect in \hei, \pab\ and \brg.  This signature might probe the existence of short-lived 
density waves forming within the inner disk beyond the corotation radius, similar to those obtained within 3D MHD simulations \citep[see Fig.~13 of][]{Romanova25b}.

\section{Summary and conclusions}
\label{sec:dis}

In this paper we presented follow-up spectropolarimetric and velocimetric observations of the cTTS TW~Hya with SPIRou at CFHT.  This new study expands a previous one from data 
recorded with the same instrument \citep{Donati24b}, adding spectra collected in 2024 and 2025 and doubling the total amount of spectra used in the analysis (now reaching 164 over 
2248~d).  
LSD was applied to all \texttt{Libre ESpRIT}-reduced spectra in a homogeneous way, to derive the Stokes $I$ and $V$ profiles used to characterize the large- and small-scale magnetic 
topology of TW~Hya and its temporal evolution.  We also derived RV and $dT$ variations with LBL from the \texttt{APERO}-reduced spectra to investigate temperature changes at the 
stellar surface and the potential presence of close-in exoplanets in the inner accretion disk.  

The longitudinal magnetic field, derived from LSD profiles, showed clear rotational modulation with a period of $3.587\pm0.009$~d as well as seasonal variations.  The latter two 
seasons in particular, each covering several SPIRou runs, indicated that the large-scale field was undergoing fluctuations on a timescale of order a year, in addition to the longer 
term variations already outlined in the previous study.  
The ZDI modeling of the LSD Stokes $I$ and $V$ profiles showed that the large-scale field mostly consisted of a poloidal dipole inclined at $17\pm7$\degr\ to the rotation axis with a 
polar strength of $0.83\pm0.11$~kG, encompassing $75\pm5$\% of the reconstructed magnetic energy ($80\pm5$\% of the poloidal field energy).  The large-scale poloidal field also 
included a complex octupole component, whose average strength was smaller than half that of the dominant dipole.  The ZDI maps confirmed that the large-scale 
field regularly evolved, with the negative radial field region close to the visible pole splitting at times in two smaller structures (as in 2024) or vice versa (as in 2025).  The 
small-scale field inferred from ZDI maps (2.7~kG) was slightly weaker than that derived with ZeeTurbo from the Zeeman broadening of unpolarized spectral lines ($3.2\pm0.2$~kG), though 
consistent with previous literature estimates \citep[e.g.,][]{Sokal18,Lavail19,Lopez-Valdivia23}. 

Once corrected from telluric pollution inducing a correlation with BERV, RVs of TW~Hya mostly showed rotational modulation caused by activity, with a period of $3.5635\pm0.0014$~d 
and an amplitude of $32\pm11$~\ms.  No clear signal emerges from the residual RVs, whose rms reaches down to 13.5~\ms.  The corresponding upper limit we derived on the mass of a
close-in planet orbiting within the inner accretion disk ranged from 0.33 to 0.98~\mjup\ for distances 0.053 to 0.41~au (assuming a circular orbit in the stellar equatorial plane, 
inclined at 10\degr\ to the line of sight).  In particular, we were not able to confirm the existence of the candidate close-in planet at an orbital period of $\simeq$8~d suggested in our 
previous study.  Our sensitivity to putative planets rapidly degrades outward as a result of the poor sampling at longer periods, with upper limits of about 3 and 15~\mjup\ for putative 
planets at distances of 1 and 2~au, including the one that may have caused the reported gap at 1~au \citep{Andrews16} or prevented the optically thick dust to reach the inner disk \citep{Calvet02}.  

Although veiling in the spectrum of TW~Hya was small most of the time (median $r_{JH}$ and $r_K$ of 0.20$-$0.25), accretion was clearly occurring from the inner accretion disk to the stellar 
surface.  From the \hei\ IRT as well as \pab\ and \brg\ emission lines, in particular their redshifted absorption components, and their temporal variations, we showed that accretion 
was unstable and irregular in most seasons, with an accretion pattern that evolved on a timescale of a few weeks (as in 2024 and 2025), as a likely response to the fluctuating 
large-scale magnetic topology and the clumpy / inhomogeneous nature of the inner accretion disk.  From the (veiling corrected) \pab\ and \brg\ line EWs, we inferred logarithmic mass 
accretion rates using the scaling relations of \citet{Fiorellino25}, equal to $\log \Mdot = -8.33\pm0.20$~dex on average with seasonal variations ranging from $-7.95\pm0.10$ in 
2020 down to $-8.59\pm0.16$ in 2024.  The derived average magnetospheric radius \rmag\ is much smaller than the corotation radius \rcor, with $\rmag/\rcor = 0.40\pm0.06$ when using the 
analytical relation of \citet{Bessolaz08}, or $0.33\pm0.06$ according to the results of the 3D MHD simulations of \citet{Romanova25b}.  This confirmed that magnetospheric accretion onto 
TW~Hya was unstable, with tongues of disk material penetrating the magnetosphere and falling on the star at intermediate latitudes \citep{Romanova25b}.  The 2025 TESS light 
curve, whose periodogram peaked at $\simeq$6.2~d but showed no power at the rotation period, further supports this view and may suggest that, as in the simulations of \citet{Romanova25b}, 
density structures in the inner accretion disk caused the observed photometric periodicity and contributed to the accretion variability.  

Given the results of this new study and the significant improvement brought over our previous conclusions, we plan to regularly repeat similar spectropolarimetric and velocimetric 
observations of TW~Hya every few seasons in the coming years to further extend our monitoring on still longer timescales.  We will include this time not only nIR SPIRou spectra but also 
optical ESPaDOnS data to expand our set of accretion and magnetic diagnostics, and achieve an even more consistent and global modeling \citep[as in, e.g.,][]{Donati24}.  This will indeed 
soon be possible thanks to the VISION / Wenaokeao bonnette aimed at carrying out simultaneous observations with both instruments, to be 
installed at the CFHT Cassegrain focus in late 2026.  We will focus in particular on the long term evolution of the large-scale magnetic topology of TW~Hya and the associated changes 
in the accretion patterns to further document cases of unstable accretion in late phases of stellar formation and disk dissipation, trying at the same time to lower the 
upper limit on the mass of putative close-in planets that may contribute to inducing density structures in the inner disk.

\section*{Data availability}  SPIRou data used in this study are publicly available at the Canadian Astronomy Data Center (\url{https://www.cadc-ccda.hia-iha.nrc-cnrc.gc.ca}).

\begin{acknowledgements}
We thank an anonymous referee for his/her insight and suggestions on an earlier version of the manuscript.  
This work benefited from the SIMBAD CDS database at URL {\tt http://simbad.u-strasbg.fr/simbad} and the ADS system at URL {\tt https://ui.adsabs.harvard.edu}, and was 
supported by the French National Research Agency (ANR) in the framework of the ``Investissements d'Avenir'' program (ANR-15-IDEX-02) and the IRYSS (ANR-23-EDIR0001-01) project.
Our study is based on data obtained at the CFHT, operated by the CNRC (Canada), INSU/CNRS (France) and the University of Hawaii.
The authors wish to recognise and acknowledge the very significant cultural role and reverence that the summit of Maunakea has always had
within the indigenous Hawaiian community.  
\end{acknowledgements}

\bibliographystyle{aa}
\bibliography{twhya2}
\clearpage

\begin{appendix}

\section{SPIRou observations: additional material}
\label{sec:appA}

Table~\ref{tab:log} provides the observation log for the SPIRou spectra of TW~Hya.

\begin{table*}[ht!]
\caption[]{Observing log of our SPIRou observations of TW~Hya}
\centering
\resizebox{0.75\linewidth}{!}{
\begin{tabular}{ccccccccccc}
\hline
BJD        & UT date & BERV   & c / $\phi$ & t$_{\rm exp}$ & S/N   & $\sigma_P$           & \Bl\  &   LBL RV &  $r_{JH}$ / $r_K$ & EW \pab\ / \brg \\
(2459000+) &         & (\kms) &            &   (s)         & ($H$) & ($10^{-4} I_c$)      &(G)    &   (\kms) &                   & (\kms)          \\
\hline
-411.2218803 & 15 Apr 2019 & -8.403 & 0 / 0.078 & 1136.7 & 168 & 4.39 & 18$\pm$24 & 0.0115$\pm$0.0037 & 0.254 / 0.090 & 309 / 68 \\
-410.0666365 & 16 Apr 2019 & -9.173 & 0 / 0.400 & 1136.7 & 191 & 3.78 & 33$\pm$24 & -0.0432$\pm$0.0033 & 0.403 / 0.244 & 560 / 134 \\
-408.2102656 & 18 Apr 2019 & -9.559 & 0 / 0.917 & 1136.7 & 164 & 4.37 & 17$\pm$25 & 0.0106$\pm$0.0036 & 0.277 / 0.131 & 368 / 83 \\
-407.1835982 & 19 Apr 2019 & -9.997 & 1 / 0.203 & 1136.7 & 182 & 3.84 & 33$\pm$22 & 0.0049$\pm$0.0034 & 0.277 / 0.127 & 359 / 77 \\
-406.2071730 & 20 Apr 2019 & -10.302 & 1 / 0.476 & 1136.7 & 180 & 4.00 & -8$\pm$23 & -0.0296$\pm$0.0036 & 0.275 / 0.273 & 354 / 74 \\
-404.2046516 & 22 Apr 2019 & -11.033 & 2 / 0.034 & 1136.7 & 185 & 3.88 & 3$\pm$21 & -0.0027$\pm$0.0036 & 0.221 / 0.277 & 223 / 41 \\
-403.1758960 & 23 Apr 2019 & -11.465 & 2 / 0.321 & 1136.7 & 144 & 5.47 & 56$\pm$32 & -0.0255$\pm$0.0043 & 0.325 / 0.425 & 279 / 62 \\
-402.2162162 & 24 Apr 2019 & -11.715 & 2 / 0.588 & 1136.7 & 177 & 4.09 & 17$\pm$23 & -0.0307$\pm$0.0034 & 0.275 / 0.425 & 340 / 75 \\
-401.1785924 & 25 Apr 2019 & -12.164 & 2 / 0.877 & 1136.7 & 198 & 3.60 & 17$\pm$20 & -0.0007$\pm$0.0036 & 0.246 / 0.448 & 390 / 77 \\
-400.2281200 & 26 Apr 2019 & -12.383 & 3 / 0.142 & 1136.7 & 184 & 3.81 & 26$\pm$22 & 0.0440$\pm$0.0034 & 0.333 / 0.656 & 528 / 99 \\
-399.2256117 & 27 Apr 2019 & -12.733 & 3 / 0.422 & 1136.7 & 143 & 5.24 & 18$\pm$29 & -0.0047$\pm$0.0044 & 0.261 / 0.720 & 190 / 38 \\
\hline
-114.9575691 & 05 Feb 2020 & 17.635 & 82 / 0.671 & 1582.4 & 286 & 2.26 & -148$\pm$13 & 0.0399$\pm$0.0025 & 0.301 / 0.232 & 624 / 120 \\
-110.9945004 & 09 Feb 2020 & 16.516 & 83 / 0.776 & 1582.4 & 257 & 2.56 & -138$\pm$14 & -0.0006$\pm$0.0026 & 0.263 / 0.155 & 592 / 111 \\
-104.0041815 & 16 Feb 2020 & 14.243 & 85 / 0.725 & 1582.4 & 213 & 3.24 & -196$\pm$20 & 0.0049$\pm$0.0030 & 0.358 / 0.317 & 632 / 127 \\
-102.9682410 & 17 Feb 2020 & 13.803 & 86 / 0.014 & 1582.4 & 252 & 2.58 & -133$\pm$15 & -0.0287$\pm$0.0027 & 0.328 / 0.416 & 623 / 126 \\
-102.0330761 & 18 Feb 2020 & 13.620 & 86 / 0.275 & 1582.4 & 267 & 2.39 & -97$\pm$14 & -0.0269$\pm$0.0026 & 0.330 / 0.424 & 604 / 116 \\
-101.0329883 & 19 Feb 2020 & 13.263 & 86 / 0.553 & 1582.4 & 285 & 2.23 & -177$\pm$13 & 0.0220$\pm$0.0025 & 0.302 / 0.314 & 653 / 126 \\
-79.0917046 & 12 Mar 2020 & 4.835 & 92 / 0.670 & 1582.4 & 287 & 2.40 & -207$\pm$13 & 0.0156$\pm$0.0025 & 0.255 / 0.246 & 516 / 101 \\
-22.2151786 & 08 May 2020 & -16.468 & 108 / 0.527 & 1582.4 & 260 & 2.79 & -186$\pm$16 & 0.0219$\pm$0.0026 & 0.302 / 0.035 & 525 / 104 \\
-21.1748547 & 09 May 2020 & -16.852 & 108 / 0.817 & 1582.4 & 275 & 2.43 & -212$\pm$14 & -0.0172$\pm$0.0025 & 0.308 / 0.039 & 428 / 85 \\
-20.2194204 & 10 May 2020 & -17.023 & 109 / 0.083 & 2005.9 & 333 & 2.00 & -143$\pm$11 & -0.0348$\pm$0.0023 & 0.289 / 0.155 & 411 / 80 \\
-19.1862407 & 11 May 2020 & -17.381 & 109 / 0.371 & 2005.9 & 335 & 2.04 & -135$\pm$11 & 0.0086$\pm$0.0024 & 0.190 / 0.123 & 319 / 58 \\
-18.1874428 & 12 May 2020 & -17.649 & 109 / 0.649 & 2005.9 & 315 & 2.17 & -183$\pm$13 & 0.0241$\pm$0.0023 & 0.322 / 0.133 & 525 / 109 \\
-17.1830447 & 13 May 2020 & -17.925 & 109 / 0.929 & 2005.9 & 305 & 2.38 & -188$\pm$14 & 0.0009$\pm$0.0026 & 0.293 / 0.100 & 554 / 111 \\
-15.1805270 & 15 May 2020 & -18.448 & 110 / 0.488 & 2005.9 & 281 & 2.45 & -192$\pm$14 & 0.0339$\pm$0.0026 & 0.287 / 0.000 & 345 / 60 \\
\hline
267.0324755 & 21 Feb 2021 & 12.079 & 189 / 0.164 & 2005.9 & 318 & 2.21 & -107$\pm$12 & 0.0369$\pm$0.0025 & 0.220 / 0.206 & 140 / 26 \\
267.9869082 & 22 Feb 2021 & 11.826 & 189 / 0.430 & 2005.9 & 346 & 1.95 & -120$\pm$10 & 0.0077$\pm$0.0023 & 0.161 / 0.125 & 140 / 25 \\
269.0011213 & 23 Feb 2021 & 11.417 & 189 / 0.713 & 2005.9 & 278 & 2.48 & -71$\pm$15 & -0.0493$\pm$0.0024 & 0.400 / 0.324 & 254 / 54 \\
271.9683085 & 26 Feb 2021 & 10.368 & 190 / 0.540 & 2005.9 & 379 & 1.77 & -84$\pm$10 & -0.0407$\pm$0.0020 & 0.270 / 0.200 & 268 / 52 \\
274.0167096 & 28 Feb 2021 & 9.474 & 191 / 0.111 & 2005.9 & 325 & 2.18 & -78$\pm$12 & 0.0344$\pm$0.0022 & 0.239 / 0.137 & 282 / 52 \\
276.0307067 & 02 Mar 2021 & 8.662 & 191 / 0.673 & 2005.9 & 246 & 3.51 & -91$\pm$20 & -0.0241$\pm$0.0029 & 0.266 / 0.138 & 179 / 29 \\
276.9398811 & 03 Mar 2021 & 8.502 & 191 / 0.926 & 2005.9 & 338 & 1.86 & -112$\pm$10 & 0.0196$\pm$0.0021 & 0.218 / 0.116 & 223 / 44 \\
277.9841904 & 04 Mar 2021 & 7.989 & 192 / 0.218 & 2005.9 & 392 & 1.66 & -95$\pm$9 & 0.0091$\pm$0.0022 & 0.201 / 0.127 & 145 / 22 \\
293.9200402 & 20 Mar 2021 & 1.572 & 196 / 0.660 & 2005.9 & 383 & 1.74 & -92$\pm$9 & -0.0314$\pm$0.0019 & 0.205 / 0.161 & 172 / 25 \\
294.9534459 & 21 Mar 2021 & 1.068 & 196 / 0.948 & 2005.9 & 362 & 1.81 & -85$\pm$10 & 0.0103$\pm$0.0020 & 0.214 / 0.155 & 126 / 19 \\
296.9619202 & 23 Mar 2021 & 0.213 & 197 / 0.508 & 2005.9 & 392 & 1.67 & -83$\pm$11 & & 0.470 / 0.493 & 379 / 91 \\
297.8864493 & 24 Mar 2021 & -0.005 & 197 / 0.766 & 2005.9 & 401 & 1.62 & -57$\pm$10 & 0.0380$\pm$0.0021 & 0.394 / 0.464 & 268 / 59 \\
299.9019941 & 26 Mar 2021 & -0.876 & 198 / 0.328 & 2005.9 & 363 & 1.78 & -89$\pm$10 & -0.0049$\pm$0.0019 & 0.281 / 0.287 & 208 / 44 \\
300.8485035 & 27 Mar 2021 & -1.147 & 198 / 0.592 & 2005.9 & 188 & 3.81 & -100$\pm$22 & -0.0228$\pm$0.0033 & 0.310 / 0.242 & 165 / 33 \\
301.8981152 & 28 Mar 2021 & -1.692 & 198 / 0.884 & 2005.9 & 190 & 4.52 & -80$\pm$26 & 0.0067$\pm$0.0029 & 0.298 / 0.522 & 121 / 21 \\
301.9183606 & 28 Mar 2021 & -1.745 & 198 / 0.890 & 2005.9 & 187 & 4.33 & -78$\pm$25 & & 0.311 / 0.466 & 125 / 19 \\
304.9025306 & 31 Mar 2021 & -2.935 & 199 / 0.722 & 2005.9 & 319 & 2.15 & -16$\pm$13 & -0.0357$\pm$0.0022 & 0.399 / 0.370 & 266 / 52 \\
305.9110261 & 01 Apr 2021 & -3.367 & 200 / 0.003 & 2005.9 & 367 & 1.74 & -63$\pm$10 & 0.0178$\pm$0.0019 & 0.306 / 0.321 & 261 / 58 \\
326.8395710 & 22 Apr 2021 & -11.338 & 205 / 0.838 & 2005.9 & 279 & 2.54 & -81$\pm$15 & 0.0074$\pm$0.0023 & 0.332 / 0.200 & 221 / 38 \\
327.8120414 & 23 Apr 2021 & -11.621 & 206 / 0.109 & 2005.9 & 383 & 1.77 & -96$\pm$10 & 0.0455$\pm$0.0019 & 0.327 / 0.407 & 318 / 57 \\
328.8393031 & 24 Apr 2021 & -12.042 & 206 / 0.395 & 2005.9 & 362 & 1.86 & -90$\pm$12 & -0.0628$\pm$0.0021 & 0.423 / 0.392 & 378 / 68 \\
329.8575315 & 25 Apr 2021 & -12.434 & 206 / 0.679 & 2005.9 & 347 & 1.94 & -65$\pm$11 & -0.0247$\pm$0.0019 & 0.298 / 0.165 & 242 / 41 \\
330.8156155 & 26 Apr 2021 & -12.670 & 206 / 0.946 & 2005.9 & 286 & 2.56 & -97$\pm$14 & 0.0161$\pm$0.0018 & 0.262 / 0.135 & 154 / 26 \\
330.8364802 & 26 Apr 2021 & -12.723 & 206 / 0.952 & 2005.9 & 257 & 3.44 & -64$\pm$19 & & 0.247 / 0.159 & 156 / 25 \\
331.8534125 & 27 Apr 2021 & -13.104 & 207 / 0.235 & 2005.9 & 343 & 2.00 & -108$\pm$11 & 0.0074$\pm$0.0020 & 0.284 / 0.157 & 301 / 72 \\
332.8680461 & 28 Apr 2021 & -13.474 & 207 / 0.518 & 2005.9 & 318 & 2.35 & -83$\pm$14 & -0.0472$\pm$0.0022 & 0.310 / 0.192 & 189 / 28 \\
334.8263125 & 30 Apr 2021 & -14.031 & 208 / 0.064 & 2005.9 & 398 & 1.73 & -66$\pm$11 & 0.0496$\pm$0.0022 & 0.420 / 0.572 & 253 / 39 \\
335.8083082 & 01 May 2021 & -14.309 & 208 / 0.338 & 2005.9 & 379 & 1.75 & -84$\pm$11 & -0.0468$\pm$0.0021 & 0.424 / 0.583 & 419 / 83 \\
336.8470324 & 02 May 2021 & -14.726 & 208 / 0.628 & 2005.9 & 293 & 2.38 & -80$\pm$15 & -0.0505$\pm$0.0023 & 0.403 / 0.467 & 313 / 55 \\
\hline
649.8912529 & 11 Mar 2022 & 5.483 & 295 / 0.899 & 2005.9 & 401 & 1.73 & -3$\pm$10 & -0.0349$\pm$0.0020 & 0.258 / 0.190 & 433 / 75 \\
650.9038460 & 12 Mar 2022 & 5.039 & 296 / 0.182 & 2005.9 & 380 & 1.81 & -14$\pm$11 & 0.0100$\pm$0.0019 & 0.317 / 0.214 & 356 / 66 \\
651.8779503 & 13 Mar 2022 & 4.695 & 296 / 0.453 & 2005.9 & 410 & 1.67 & -3$\pm$10 & 0.0302$\pm$0.0021 & 0.299 / 0.285 & 413 / 87 \\
652.8769485 & 14 Mar 2022 & 4.285 & 296 / 0.732 & 2005.9 & 398 & 1.70 & 9$\pm$10 & -0.0185$\pm$0.0020 & 0.284 / 0.292 & 366 / 83 \\
653.9841063 & 15 Mar 2022 & 3.591 & 297 / 0.040 & 2005.9 & 389 & 1.74 & -30$\pm$9 & -0.0473$\pm$0.0022 & 0.192 / 0.182 & 387 / 80 \\
654.9433060 & 16 Mar 2022 & 3.283 & 297 / 0.308 & 1002.9 & 248 & 2.73 & -3$\pm$16 & -0.0040$\pm$0.0024 & 0.302 / 0.209 & 409 / 88 \\
655.9057570 & 17 Mar 2022 & 2.969 & 297 / 0.576 & 2005.9 & 328 & 2.53 & 11$\pm$14 & -0.0079$\pm$0.0023 & 0.289 / 0.210 & 317 / 63 \\
657.9321715 & 19 Mar 2022 & 2.070 & 298 / 0.141 & 2005.9 & 396 & 1.74 & -39$\pm$9 & -0.0192$\pm$0.0023 & 0.176 / 0.098 & 226 / 43 \\
658.9416328 & 20 Mar 2022 & 1.630 & 298 / 0.423 & 2005.9 & 353 & 1.97 & -4$\pm$11 & 0.0272$\pm$0.0020 & 0.271 / 0.176 & 256 / 45 \\
659.9525078 & 21 Mar 2022 & 1.187 & 298 / 0.704 & 2005.9 & 288 & 2.49 & 0$\pm$15 & -0.0090$\pm$0.0023 & 0.365 / 0.117 & 385 / 79 \\
660.9330715 & 22 Mar 2022 & 0.822 & 298 / 0.978 & 2005.9 & 313 & 2.24 & 29$\pm$13 & -0.0212$\pm$0.0021 & 0.323 / 0.068 & 368 / 67 \\
661.8884647 & 23 Mar 2022 & 0.525 & 299 / 0.244 & 2005.9 & 188 & 4.54 & 11$\pm$27 & & 0.322 / 0.000 & 259 / 46 \\
661.9123263 & 23 Mar 2022 & 0.461 & 299 / 0.251 & 2005.9 & 231 & 3.51 & -15$\pm$20 & -0.0086$\pm$0.0024 & 0.288 / 0.004 & 255 / 44 \\
678.9220651 & 09 Apr 2022 & -6.539 & 303 / 0.993 & 2005.9 & 360 & 1.94 & -32$\pm$11 & -0.0124$\pm$0.0020 & 0.275 / 0.195 & 257 / 46 \\
681.8732563 & 12 Apr 2022 & -7.591 & 304 / 0.816 & 2005.9 & 293 & 2.42 & -28$\pm$15 & -0.0424$\pm$0.0023 & 0.429 / 0.237 & 326 / 54 \\
682.8548885 & 13 Apr 2022 & -7.929 & 305 / 0.089 & 2005.9 & 332 & 2.04 & -25$\pm$13 & -0.0034$\pm$0.0021 & 0.435 / 0.386 & 427 / 77 \\
683.8634512 & 14 Apr 2022 & -8.335 & 305 / 0.370 & 2005.9 & 350 & 1.95 & 14$\pm$12 & 0.0074$\pm$0.0020 & 0.402 / 0.456 & 317 / 55 \\
684.7962789 & 15 Apr 2022 & -8.539 & 305 / 0.630 & 2005.9 & 365 & 1.88 & -13$\pm$11 & -0.0274$\pm$0.0021 & 0.354 / 0.441 & 290 / 50 \\
687.8950037 & 18 Apr 2022 & -9.918 & 306 / 0.494 & 2005.9 & 344 & 2.05 & 13$\pm$13 & 0.0254$\pm$0.0022 & 0.485 / 0.413 & 340 / 55 \\
690.8404521 & 21 Apr 2022 & -10.879 & 307 / 0.315 & 2005.9 & 356 & 1.98 & 4$\pm$13 & 0.0237$\pm$0.0024 & 0.484 / 0.607 & 276 / 51 \\
710.8180850 & 11 May 2022 & -17.271 & 312 / 0.885 & 2005.9 & 263 & 2.82 & 16$\pm$17 & -0.0136$\pm$0.0025 & 0.381 / 0.139 & 255 / 41 \\
711.8161622 & 12 May 2022 & -17.537 & 313 / 0.163 & 2005.9 & 383 & 1.78 & -0$\pm$11 & 0.0364$\pm$0.0021 & 0.338 / 0.285 & 290 / 62 \\
712.8124998 & 13 May 2022 & -17.794 & 313 / 0.441 & 2005.9 & 300 & 2.36 & 16$\pm$15 & 0.0134$\pm$0.0023 & 0.390 / 0.210 & 261 / 43 \\
713.7983426 & 14 May 2022 & -18.021 & 313 / 0.716 & 2005.9 & 300 & 2.32 & 38$\pm$15 & -0.0347$\pm$0.0023 & 0.457 / 0.246 & 373 / 85 \\
714.7779533 & 15 May 2022 & -18.227 & 313 / 0.989 & 2005.9 & 349 & 1.98 & 29$\pm$12 & -0.0196$\pm$0.0022 & 0.423 / 0.416 & 324 / 56 \\
715.8090560 & 16 May 2022 & -18.552 & 314 / 0.276 & 2005.9 & 338 & 2.06 & -6$\pm$13 & 0.0042$\pm$0.0022 & 0.404 / 0.489 & 253 / 47 \\
716.8108812 & 17 May 2022 & -18.802 & 314 / 0.556 & 2005.9 & 385 & 1.80 & 15$\pm$11 & -0.0217$\pm$0.0022 & 0.371 / 0.451 & 328 / 68 \\
719.8653011 & 20 May 2022 & -19.617 & 315 / 0.407 & 2005.9 & 342 & 2.11 & -11$\pm$13 & -0.0106$\pm$0.0046 & 0.375 / 0.605 & 217 / 31 \\
\hline
\end{tabular}
}
\tablefoot{For each visit, we list the barycentric Julian date BJD, the UT date, the barycentric Earth RV (BERV), the rotation cycle c and phase $\phi$ (computed
as indicated in Sec.~\ref{sec:obs}), the total observing time t$_{\rm exp}$, the peak S/N in the spectrum (in the $H$ band) per 2.3~\kms\ pixel, the noise level
in the LSD Stokes $V$ profile, the BERV-corrected LBL RVs and error bar (see Sec.~\ref{sec:rvs}), the estimated \Bl\ from LSD profiles of atomic lines (with error bars), 
the measured veiling in the $JH$ and $K$ bands, $r_{JH}$ and $r_K$, and finally the measured \pab\ and \brg\ EWs (uncorrected for veiling). }
\label{tab:log}
\end{table*}

\setcounter{table}{0}
\begin{table*}[ht!]
\caption[]{(continued)}
\centering
\resizebox{0.75\linewidth}{!}{
\begin{tabular}{ccccccccccc}
\hline
BJD        & UT date & BERV   & c / $\phi$ & t$_{\rm exp}$ & S/N   & $\sigma_P$           & \Bl\  &   LBL RV &  $r_{JH}$ / $r_K$ & EW \pab\ / \brg \\
(2459000+) &         & (\kms) &            &   (s)         & ($H$) & ($10^{-4} I_c$)      &(G)    &   (\kms) &                   & (\kms)          \\
\hline
1355.9755282 & 15 Feb 2024 & 14.617 & 492 / 0.745 & 2005.9 & 370 & 1.70 & -23$\pm$9 & -0.0391$\pm$0.0020 & 0.148 / 0.299 & 179 / 28 \\
1358.9832474 & 18 Feb 2024 & 13.550 & 493 / 0.583 & 2005.9 & 345 & 1.92 & -64$\pm$9 & -0.0039$\pm$0.0021 & 0.111 / 0.061 & 147 / 22 \\
1360.0249145 & 19 Feb 2024 & 13.086 & 493 / 0.874 & 2005.9 & 297 & 2.13 & -21$\pm$11 & 0.0117$\pm$0.0022 & 0.172 / 0.066 & 214 / 35 \\
1360.9607386 & 20 Feb 2024 & 12.893 & 494 / 0.135 & 2005.9 & 316 & 2.03 & -55$\pm$11 & 0.0452$\pm$0.0021 & 0.179 / 0.116 & 188 / 31 \\
1361.9989955 & 21 Feb 2024 & 12.431 & 494 / 0.424 & 2005.9 & 363 & 1.77 & -71$\pm$9 & -0.0001$\pm$0.0020 & 0.133 / 0.195 & 181 / 31 \\
1362.9445027 & 22 Feb 2024 & 12.207 & 494 / 0.688 & 2005.9 & 375 & 1.71 & -39$\pm$9 & -0.0163$\pm$0.0020 & 0.148 / 0.217 & 124 / 20 \\
1363.9867042 & 23 Feb 2024 & 11.728 & 494 / 0.978 & 2005.9 & 356 & 1.84 & -52$\pm$10 & 0.0345$\pm$0.0021 & 0.249 / 0.317 & 260 / 51 \\
1364.9998426 & 24 Feb 2024 & 11.322 & 495 / 0.261 & 2005.9 & 371 & 1.76 & -70$\pm$9 & 0.0186$\pm$0.0019 & 0.149 / 0.226 & 200 / 32 \\
1367.0023952 & 26 Feb 2024 & 10.562 & 495 / 0.819 & 2005.9 & 340 & 1.92 & -57$\pm$10 & 0.0005$\pm$0.0019 & 0.203 / 0.140 & 248 / 48 \\
1367.9784327 & 27 Feb 2024 & 10.243 & 496 / 0.091 & 2005.9 & 347 & 1.92 & -58$\pm$10 & 0.0465$\pm$0.0020 & 0.158 / 0.066 & 171 / 36 \\
1369.0106120 & 28 Feb 2024 & 9.776 & 496 / 0.379 & 2005.9 & 350 & 1.89 & -91$\pm$9 & 0.0047$\pm$0.0021 & 0.114 / 0.051 & 110 / 19 \\
1369.9360077 & 29 Feb 2024 & 9.585 & 496 / 0.637 & 2005.9 & 337 & 2.02 & -83$\pm$10 & 0.0025$\pm$0.0022 & 0.157 / 0.066 & 107 / 16 \\
1370.9693403 & 01 Mar 2024 & 9.107 & 496 / 0.925 & 2005.9 & 195 & 3.69 & -94$\pm$21 & 0.0389$\pm$0.0021 & 0.264 / 0.101 & 191 / 38 \\
1370.9875473 & 01 Mar 2024 & 9.060 & 496 / 0.930 & 1002.9 & 229 & 2.87 & -67$\pm$15 & & 0.212 / 0.096 & 186 / 34 \\
1371.9079342 & 02 Mar 2024 & 8.877 & 497 / 0.186 & 2005.9 & 359 & 1.83 & -84$\pm$10 & 0.0273$\pm$0.0019 & 0.195 / 0.121 & 142 / 23 \\
1385.9468478 & 16 Mar 2024 & 3.056 & 501 / 0.100 & 2005.9 & 257 & 2.55 & -82$\pm$14 & 0.0182$\pm$0.0024 & 0.238 / 0.303 & 284 / 67 \\
1386.9425836 & 17 Mar 2024 & 2.650 & 501 / 0.378 & 2005.9 & 376 & 1.68 & -101$\pm$9 & -0.0093$\pm$0.0019 & 0.193 / 0.290 & 262 / 55 \\
1388.9288974 & 19 Mar 2024 & 1.854 & 501 / 0.932 & 2005.9 & 387 & 1.63 & -79$\pm$9 & 0.0095$\pm$0.0019 & 0.204 / 0.337 & 229 / 42 \\
1390.9747852 & 21 Mar 2024 & 0.905 & 502 / 0.502 & 2005.9 & 331 & 2.02 & -108$\pm$11 & -0.0278$\pm$0.0020 & 0.220 / 0.212 & 131 / 22 \\
1391.9167761 & 22 Mar 2024 & 0.638 & 502 / 0.765 & 2005.9 & 313 & 2.06 & -82$\pm$11 & 0.0022$\pm$0.0020 & 0.193 / 0.159 & 81 / 14 \\
1392.9174745 & 23 Mar 2024 & 0.221 & 503 / 0.044 & 2005.9 & 348 & 1.81 & -84$\pm$9 & 0.0428$\pm$0.0019 & 0.159 / 0.149 & 83 / 15 \\
1398.9301107 & 29 Mar 2024 & -2.293 & 504 / 0.720 & 2005.9 & 346 & 1.89 & -53$\pm$9 & 0.0003$\pm$0.0020 & 0.132 / 0.169 & 144 / 25 \\
1399.9262821 & 30 Mar 2024 & -2.694 & 504 / 0.998 & 2005.9 & 356 & 1.75 & -90$\pm$9 & 0.0147$\pm$0.0020 & 0.112 / 0.129 & 160 / 27 \\
1401.9026360 & 01 Apr 2024 & -3.453 & 505 / 0.549 & 2005.9 & 367 & 1.75 & -113$\pm$8 & -0.0064$\pm$0.0022 & 0.097 / 0.090 & 58 / 9 \\
1402.8901039 & 02 Apr 2024 & -3.829 & 505 / 0.824 & 2005.9 & 365 & 1.72 & -70$\pm$8 & 0.0082$\pm$0.0020 & 0.101 / 0.117 & 68 / 11 \\
1404.9310577 & 04 Apr 2024 & -4.749 & 506 / 0.393 & 2005.9 & 337 & 1.96 & -89$\pm$10 & -0.0107$\pm$0.0020 & 0.207 / 0.205 & 182 / 28 \\
1421.8339450 & 21 Apr 2024 & -11.058 & 511 / 0.105 & 2005.9 & 358 & 1.78 & -117$\pm$9 & 0.0223$\pm$0.0019 & 0.119 / 0.154 & 144 / 30 \\
1422.8352161 & 22 Apr 2024 & -11.419 & 511 / 0.384 & 2005.9 & 342 & 1.91 & -135$\pm$9 & -0.0008$\pm$0.0022 & 0.114 / 0.141 & 98 / 15 \\
1423.8395213 & 23 Apr 2024 & -11.783 & 511 / 0.664 & 2005.9 & 370 & 1.77 & -136$\pm$9 & -0.0037$\pm$0.0020 & 0.152 / 0.167 & 166 / 35 \\
1424.8052787 & 24 Apr 2024 & -12.044 & 511 / 0.933 & 2005.9 & 392 & 1.57 & -120$\pm$9 & 0.0207$\pm$0.0018 & 0.263 / 0.286 & 315 / 81 \\
1425.8455689 & 25 Apr 2024 & -12.494 & 512 / 0.223 & 2005.9 & 359 & 1.81 & -121$\pm$10 & 0.0104$\pm$0.0019 & 0.185 / 0.221 & 183 / 34 \\
1426.8037199 & 26 Apr 2024 & -12.729 & 512 / 0.491 & 2005.9 & 360 & 1.81 & -107$\pm$9 & -0.0294$\pm$0.0019 & 0.157 / 0.138 & 123 / 23 \\
1427.8194178 & 27 Apr 2024 & -13.109 & 512 / 0.774 & 2005.9 & 377 & 1.71 & -106$\pm$9 & -0.0018$\pm$0.0020 & 0.156 / 0.161 & 144 / 25 \\
1428.8607862 & 28 Apr 2024 & -13.548 & 513 / 0.064 & 2005.9 & 358 & 1.86 & -135$\pm$10 & 0.0241$\pm$0.0020 & 0.179 / 0.225 & 247 / 48 \\
1429.8470320 & 29 Apr 2024 & -13.846 & 513 / 0.339 & 2005.9 & 275 & 2.45 & -115$\pm$14 & -0.0053$\pm$0.0024 & 0.273 / 0.212 & 170 / 31 \\
1431.8553355 & 01 May 2024 & -14.518 & 513 / 0.899 & 2005.9 & 353 & 1.87 & -132$\pm$9 & 0.0214$\pm$0.0020 & 0.143 / 0.099 & 173 / 39 \\
1432.8485753 & 02 May 2024 & -14.821 & 514 / 0.176 & 2005.9 & 385 & 1.73 & -133$\pm$8 & 0.0246$\pm$0.0022 & 0.098 / 0.116 & 119 / 20 \\
1433.7736095 & 03 May 2024 & -14.949 & 514 / 0.434 & 2005.9 & 377 & 1.77 & -149$\pm$9 & -0.0069$\pm$0.0021 & 0.136 / 0.162 & 239 / 58 \\
\hline
1713.0158448 & 06 Feb 2025 & 17.170 & 592 / 0.282 & 2005.9 & 319 & 2.06 & -69$\pm$11 & 0.0159$\pm$0.0021 & 0.189 / 0.194 & 137 / 23 \\
1714.0093731 & 07 Feb 2025 & 16.881 & 592 / 0.559 & 2005.9 & 315 & 2.09 & -73$\pm$12 & 0.0756$\pm$0.0022 & 0.287 / 0.256 & 223 / 45 \\
1715.0057602 & 08 Feb 2025 & 16.579 & 592 / 0.837 & 2005.9 & 272 & 2.52 & -94$\pm$13 & -0.0486$\pm$0.0023 & 0.196 / 0.126 & 122 / 20 \\
1715.9753874 & 09 Feb 2025 & 16.342 & 593 / 0.107 & 2005.9 & 276 & 2.75 & -84$\pm$15 & -0.0786$\pm$0.0025 & 0.197 / 0.174 & 233 / 52 \\
1718.0029573 & 11 Feb 2025 & 15.628 & 593 / 0.672 & 2005.9 & 388 & 1.74 & -98$\pm$10 & 0.0123$\pm$0.0022 & 0.262 / 0.331 & 301 / 66 \\
1719.0077673 & 12 Feb 2025 & 15.288 & 593 / 0.953 & 2005.9 & 375 & 1.77 & -117$\pm$9 & -0.0424$\pm$0.0020 & 0.124 / 0.159 & 121 / 18 \\
1719.9910647 & 13 Feb 2025 & 14.999 & 594 / 0.227 & 2005.9 & 387 & 1.78 & -49$\pm$9 & -0.0107$\pm$0.0021 & 0.121 / 0.193 & 95 / 15 \\
1720.9832593 & 14 Feb 2025 & 14.683 & 594 / 0.503 & 2005.9 & 370 & 1.71 & -86$\pm$9 & 0.0268$\pm$0.0019 & 0.134 / 0.226 & 124 / 21 \\
1722.0141227 & 15 Feb 2025 & 14.263 & 594 / 0.791 & 2005.9 & 367 & 1.76 & -101$\pm$9 & -0.0308$\pm$0.0019 & 0.152 / 0.222 & 119 / 17 \\
1723.0106466 & 16 Feb 2025 & 13.927 & 595 / 0.068 & 2005.9 & 383 & 1.77 & -44$\pm$10 & -0.0379$\pm$0.0021 & 0.238 / 0.307 & 168 / 30 \\
1724.0296283 & 17 Feb 2025 & 13.530 & 595 / 0.353 & 2005.9 & 370 & 2.09 & -74$\pm$11 & 0.0289$\pm$0.0021 & 0.208 / 0.328 & 160 / 25 \\
1725.0007129 & 18 Feb 2025 & 13.256 & 595 / 0.623 & 2005.9 & 395 & 1.72 & -83$\pm$9 & 0.0196$\pm$0.0021 & 0.211 / 0.337 & 173 / 27 \\
1726.0442307 & 19 Feb 2025 & 12.784 & 595 / 0.914 & 2005.9 & 359 & 1.88 & -79$\pm$11 & -0.0821$\pm$0.0025 & 0.339 / 0.436 & 313 / 81 \\
1746.9771950 & 12 Mar 2025 & 4.740 & 601 / 0.750 & 2005.9 & 281 & 2.32 & -102$\pm$12 & -0.0158$\pm$0.0024 & 0.151 / 0.038 & 288 / 61 \\
1747.9170765 & 13 Mar 2025 & 4.487 & 602 / 0.012 & 2005.9 & 227 & 2.93 & -107$\pm$15 & -0.0160$\pm$0.0028 & 0.158 / 0.043 & 342 / 74 \\
1748.9365046 & 14 Mar 2025 & 4.024 & 602 / 0.296 & 2005.9 & 182 & 3.72 & -55$\pm$20 & 0.0457$\pm$0.0034 & 0.237 / 0.000 & 406 / 86 \\
1751.9068854 & 17 Mar 2025 & 2.863 & 603 / 0.124 & 2005.9 & 362 & 1.90 & -82$\pm$10 & -0.0169$\pm$0.0020 & 0.167 / 0.155 & 437 / 96 \\
1752.9390149 & 18 Mar 2025 & 2.363 & 603 / 0.412 & 2005.9 & 269 & 2.79 & -58$\pm$15 & 0.0346$\pm$0.0025 & 0.216 / 0.178 & 457 / 100 \\
1755.9381703 & 21 Mar 2025 & 1.119 & 604 / 0.248 & 2005.9 & 182 & 6.44 & -100$\pm$36 & 0.0347$\pm$0.0038 & 0.250 / 0.402 & 267 / 47 \\
1775.8169239 & 10 Apr 2025 & -6.759 & 609 / 0.790 & 2005.9 & 404 & 1.64 & -131$\pm$8 & -0.0256$\pm$0.0019 & 0.135 / 0.174 & 407 / 82 \\
1776.8551860 & 11 Apr 2025 & -7.251 & 610 / 0.080 & 2005.9 & 298 & 2.29 & -83$\pm$13 & 0.0297$\pm$0.0022 & 0.266 / 0.240 & 438 / 102 \\
1777.8478327 & 12 Apr 2025 & -7.619 & 610 / 0.356 & 2005.9 & 224 & 3.24 & -82$\pm$17 & 0.0519$\pm$0.0028 & 0.212 / 0.163 & 240 / 46 \\
1778.8928299 & 13 Apr 2025 & -8.121 & 610 / 0.648 & 2005.9 & 378 & 1.71 & -115$\pm$9 & -0.0069$\pm$0.0019 & 0.160 / 0.181 & 369 / 67 \\
1779.8492038 & 14 Apr 2025 & -8.391 & 610 / 0.914 & 2005.9 & 405 & 1.63 & -107$\pm$9 & -0.0214$\pm$0.0019 & 0.198 / 0.268 & 552 / 128 \\
1780.8926822 & 15 Apr 2025 & -8.883 & 611 / 0.205 & 2005.9 & 389 & 1.70 & -76$\pm$9 & 0.0348$\pm$0.0019 & 0.211 / 0.258 & 472 / 91 \\
1781.8830603 & 16 Apr 2025 & -9.236 & 611 / 0.481 & 2005.9 & 368 & 1.74 & -87$\pm$9 & 0.0370$\pm$0.0019 & 0.183 / 0.225 & 369 / 72 \\
1802.8179611 & 07 May 2025 & -16.209 & 617 / 0.318 & 2005.9 & 338 & 1.98 & -124$\pm$10 & 0.0300$\pm$0.0021 & 0.124 / 0.161 & 263 / 50 \\
1803.8255661 & 08 May 2025 & -16.517 & 617 / 0.598 & 2005.9 & 372 & 1.76 & -141$\pm$9 & -0.0019$\pm$0.0020 & 0.138 / 0.162 & 357 / 70 \\
1804.8180791 & 09 May 2025 & -16.784 & 617 / 0.875 & 2005.9 & 362 & 1.84 & -144$\pm$9 & -0.0208$\pm$0.0021 & 0.133 / 0.084 & 196 / 34 \\
1806.8243980 & 11 May 2025 & -17.354 & 618 / 0.434 & 2005.9 & 381 & 1.72 & -132$\pm$9 & 0.0173$\pm$0.0020 & 0.156 / 0.233 & 418 / 88 \\
1807.7995671 & 12 May 2025 & -17.565 & 618 / 0.706 & 2005.9 & 360 & 1.80 & -135$\pm$9 & -0.0285$\pm$0.0020 & 0.151 / 0.251 & 427 / 97 \\
1808.7918429 & 13 May 2025 & -17.812 & 618 / 0.983 & 2005.9 & 374 & 1.71 & -129$\pm$9 & -0.0083$\pm$0.0020 & 0.180 / 0.281 & 356 / 74 \\
1809.7939048 & 14 May 2025 & -18.077 & 619 / 0.262 & 2005.9 & 326 & 1.96 & -116$\pm$11 & 0.0378$\pm$0.0021 & 0.218 / 0.446 & 349 / 63 \\
1810.8038378 & 15 May 2025 & -18.357 & 619 / 0.544 & 2005.9 & 237 & 3.02 & -149$\pm$17 & 0.0023$\pm$0.0028 & 0.292 / 0.574 & 360 / 69 \\
1812.7888575 & 17 May 2025 & -18.818 & 620 / 0.097 & 2005.9 & 378 & 1.76 & -148$\pm$9 & 0.0142$\pm$0.0019 & 0.159 / 0.183 & 300 / 56 \\
1813.8009631 & 18 May 2025 & -19.087 & 620 / 0.379 & 2005.9 & 403 & 1.66 & -118$\pm$9 & 0.0388$\pm$0.0019 & 0.230 / 0.285 & 351 / 69 \\
1814.7867835 & 19 May 2025 & -19.289 & 620 / 0.654 & 2005.9 & 399 & 1.68 & -152$\pm$9 & -0.0368$\pm$0.0020 & 0.189 / 0.237 & 305 / 62 \\
1830.7502302 & 04 Jun 2025 & -22.171 & 625 / 0.105 & 2005.9 & 280 & 2.52 & -166$\pm$14 & 0.0039$\pm$0.0025 & 0.226 / 0.105 & 269 / 45 \\
1831.7455728 & 05 Jun 2025 & -22.291 & 625 / 0.382 & 2005.9 & 342 & 2.07 & -174$\pm$11 & 0.0153$\pm$0.0023 & 0.202 / 0.137 & 319 / 67 \\
1832.7466103 & 06 Jun 2025 & -22.418 & 625 / 0.661 & 2005.9 & 354 & 1.98 & -202$\pm$10 & -0.0204$\pm$0.0022 & 0.118 / 0.156 & 193 / 30 \\
1833.7450919 & 07 Jun 2025 & -22.533 & 625 / 0.940 & 2005.9 & 361 & 1.91 & -157$\pm$10 & -0.0042$\pm$0.0021 & 0.189 / 0.210 & 295 / 60 \\
1834.7505893 & 08 Jun 2025 & -22.657 & 626 / 0.220 & 2005.9 & 357 & 1.95 & -139$\pm$11 & 0.0418$\pm$0.0022 & 0.213 / 0.210 & 292 / 58 \\
1835.7469814 & 09 Jun 2025 & -22.755 & 626 / 0.498 & 2005.9 & 380 & 1.80 & -187$\pm$9 & 0.0145$\pm$0.0020 & 0.178 / 0.139 & 332 / 61 \\
1836.7401298 & 10 Jun 2025 & -22.839 & 626 / 0.774 & 1002.9 & 274 & 2.52 & -179$\pm$14 & -0.0389$\pm$0.0023 & 0.255 / 0.167 & 412 / 77 \\
\hline
\end{tabular}
}
\end{table*}

\section{Magnetic field and temperature changes: Additional material}

\begin{figure*}[ht!]
\centerline{\includegraphics[scale=0.6,bb=20 40 900 610]{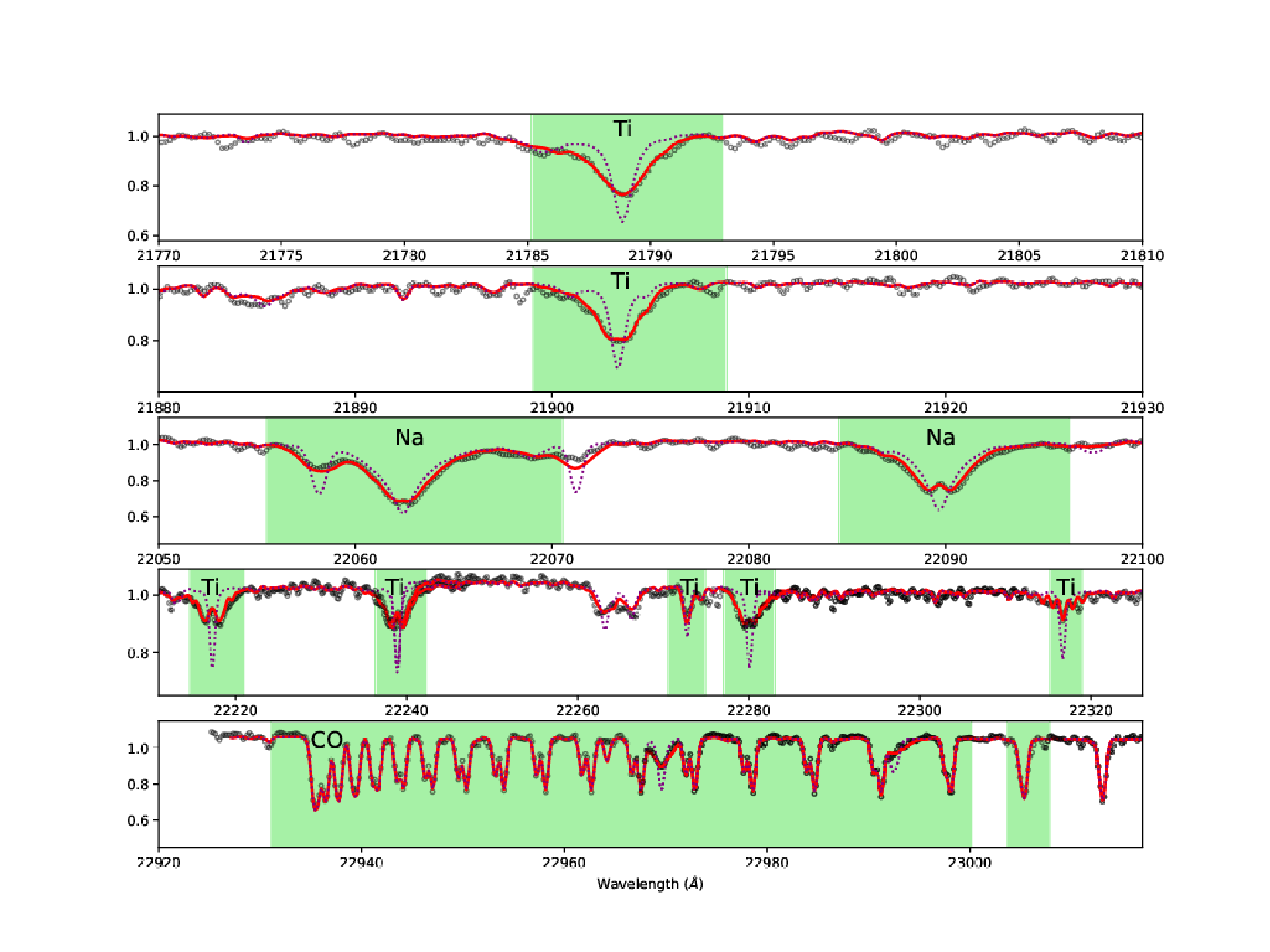}}
\caption[]{Small portion of our template SPIRou spectrum of TW~Hya in the $K$ band (black circles), along with the optimal fit achieved
with ZeeTurbo (red line) using the modeling approach of \citet{Cristofari23}.  The green 
areas indicate spectral regions considered in the fit, and the purple dotted line shows the model with no magnetic field. }
\label{fig:spc}
\end{figure*}

Figure~\ref{fig:spc} shows the fit with ZeeTurbo to the template SPIRou spectrum of TW~Hya in the $K$ band.  

\section{ZDI modeling: Additional material}
\label{sec:appC}

\begin{figure*}[ht!]
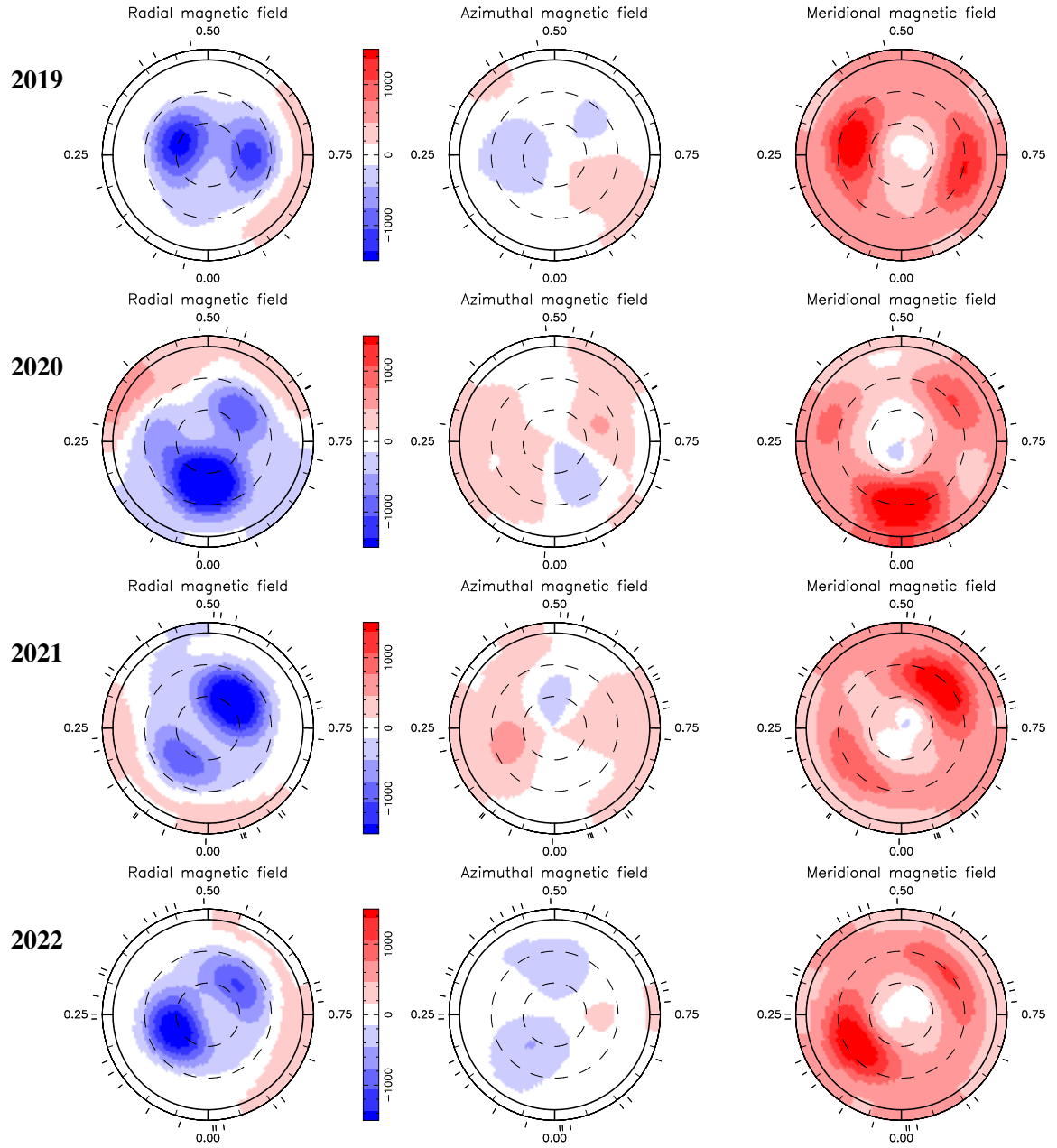
 
\centerline{\large\bf 2019\raisebox{0.30\totalheight}{\includegraphics[scale=0.40,angle=-90]{fig2/twhya2-map19.ps}}\vspace{1mm}}
\centerline{\large\bf 2020\raisebox{0.30\totalheight}{\includegraphics[scale=0.40,angle=-90]{fig2/twhya2-map20.ps}}\vspace{1mm}}
\centerline{\large\bf 2021\raisebox{0.30\totalheight}{\includegraphics[scale=0.40,angle=-90]{fig2/twhya2-map21.ps}}\vspace{1mm}}
\centerline{\large\bf 2022\raisebox{0.30\totalheight}{\includegraphics[scale=0.40,angle=-90]{fig2/twhya2-map22.ps}}} 
\caption[]{Same as Fig.~\ref{fig:map} for seasons 2019 to 2022} 
\label{fig:map2}
\end{figure*}

Figure~\ref{fig:map2} shows the reconstructed magnetic maps for seasons 2019 to 2022.  

\section{Accretion and activity: Additional material}
\label{sec:appE}

Figure~\ref{fig:brg} shows the 2D periodogram of the \brg\ line in seasons 2024 and 2025.  Figures~\ref{fig:hei2} and \ref{fig:pab2} show similar plots for 2025a 
and 2025b, for the \hei\ and \pab\ lines.  Fig.~\ref{fig:tes} shows the 2025 TESS light curve of TW~Hya and the corresponding stacked periodogram.  

\begin{figure*}[ht!]
\resizebox{0.95\textwidth}{!}{
\centerline{\hspace{-2mm}\includegraphics[scale=0.32,angle=-90]{fig2/twhya2-brg24.ps}\hspace{14.5mm}\includegraphics[scale=0.32,angle=-90]{fig2/twhya2-brg25.ps}\vspace{2mm}}}
\resizebox{0.95\textwidth}{!}{
\centerline{\includegraphics[scale=0.55,angle=-90]{fig2/twhya2-brg24per.ps}\hspace{3mm}\includegraphics[scale=0.55,angle=-90]{fig2/twhya2-brg25per.ps}}}
\caption[]{
\emr Same as Fig.~\ref{fig:hei} for the 2268~nm \brg\ line (left: 2024, right: 2025).}
\label{fig:brg}
\end{figure*}

\begin{figure*}[ht!]
\resizebox{0.95\textwidth}{!}{
\centerline{\hspace{-2mm}\includegraphics[scale=0.32,angle=-90]{fig2/twhya2-hei25a.ps}\hspace{14.5mm}\includegraphics[scale=0.32,angle=-90]{fig2/twhya2-hei25b.ps}\vspace{2mm}}}
\resizebox{0.95\textwidth}{!}{
\centerline{\includegraphics[scale=0.55,angle=-90]{fig2/twhya2-hei25aper-2.ps}\hspace{3mm}\includegraphics[scale=0.55,angle=-90]{fig2/twhya2-hei25bper-2.ps}}}
\caption[]{
Same as Fig.~\ref{fig:hei} for 2025a (left panel) and 2025b (right panel).}
\label{fig:hei2}
\end{figure*}

\begin{figure*}[ht!]
\resizebox{0.95\textwidth}{!}{
\centerline{\hspace{-2mm}\includegraphics[scale=0.32,angle=-90]{fig2/twhya2-pab25a.ps}\hspace{14.5mm}\includegraphics[scale=0.32,angle=-90]{fig2/twhya2-pab25b.ps}\vspace{2mm}}}
\resizebox{0.95\textwidth}{!}{
\centerline{\includegraphics[scale=0.55,angle=-90]{fig2/twhya2-pab25aper.ps}\hspace{3mm}\includegraphics[scale=0.55,angle=-90]{fig2/twhya2-pab25bper.ps}}}
\caption[]{
Same as Fig.~\ref{fig:pab} for 2025a (left panel) and 2025b (right panel).}
\label{fig:pab2}
\end{figure*}

\begin{figure*}
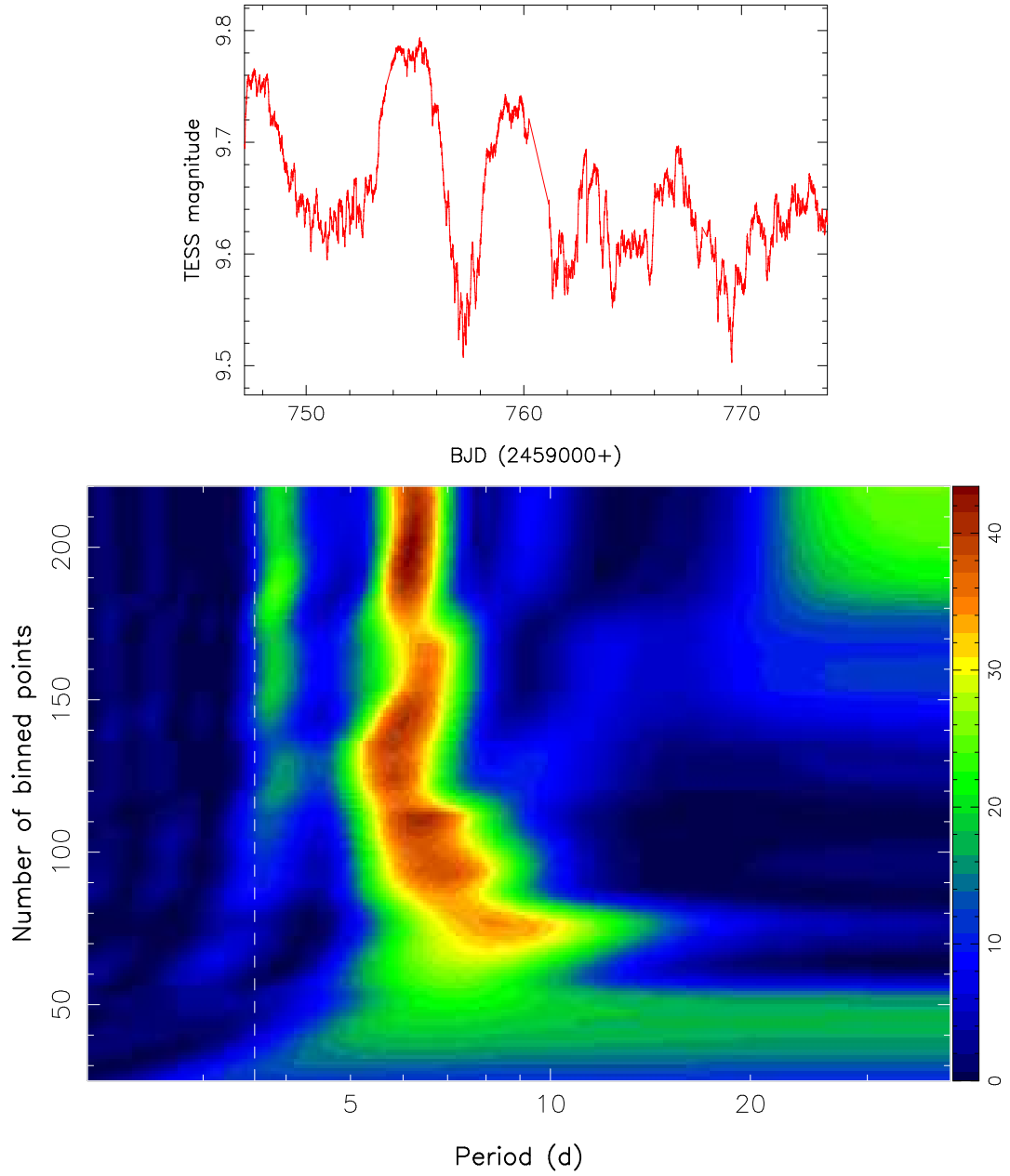

\centerline{\includegraphics[scale=0.4,angle=-90]{fig2/twhya2-tesslc.ps}\vspace{2mm}}
\centerline{\includegraphics[scale=0.5,angle=-90]{fig2/twhya2-tess25.ps}}
\caption[]{TESS light curve of TW~Hya in 2025 (March 12 to April 08, sector 90).  Top panel: light curve derived with {\tt TESSExtractor} \citep{Serna21}.  Bottom panel: stacked periodogram, 
with the TESS data binned by groups of 50 adjacent points.  In this plot, each horizontal line corresponds to a color-coded periodogram, computed on an increasing number of binned points 
(starting from the first binned point).  The color scale depicts the logarithmic power of the periodogram.  The vertical dashed line depicts the rotation period.  }
\label{fig:tes}
\end{figure*}

\FloatBarrier 

\end{appendix}
\end{document}